\documentclass[atmosphere,reprint,superscriptaddress,twocolumn,showpacs,noeprint]{revtex4-1}
\usepackage[pdftex]{graphicx}
\usepackage{braket}
\usepackage{natbib}
\usepackage{here}
\usepackage{dcolumn}
\usepackage{color}
\usepackage{bm}
\usepackage[version=3]{mhchem}
\usepackage{xspace}
\newcommand{\AffiliationISSP}{\affiliation{Institute for Solid State Physics, University of Tokyo, Kashiwanoha, Chiba 277-8581, Japan}}
\newcommand{\AffiliationUTPhys}{\affiliation{Department of Physics, University of Tokyo, Hongo, Tokyo 113-0033, Japan}}
\newcommand{\AffiliationHZB}{\affiliation{Helmholtz Zentrum-Berlin f\"{u}r Materialien und Energie GmbH, Albert-Einstein-Stra\ss e 15, 12489 Berlin, Germany}}
\newcommand{\AffiliationTitech}{\affiliation{Laboratry for Materials and Structures, Tokyo Institute of Technology, 4259 Nagatsuta-cho, Midori-ku, Yokohama 226-8503, Japan}} 
\newcommand{\AffiliationPF}{\affiliation{Photon Factory, Institute of Materials Structure Science, High Energy Accelerator Research Organization (KEK), 1-1 Oho, Tsukuba 305-0801, Japan}}
\newcommand{\SFO}{\ce{SrFeO_{3$-\delta$}}\xspace}
\newcommand{\LSFO}{\ce{La1/3Sr2/3FeO3}\xspace}
\usepackage{amsmath} 
\begin{document}
\title{Photoinduced Transient States of Antiferromagnetic Orderings in \LSFO and \SFO Thin Films Observed through Time-resolved Resonant Soft X-ray Scattering}
\author{Kohei~Yamamoto}
\email{yamako@ims.ac.jp}
\homepage{https://sites.google.com/site/yamakolux/}
\AffiliationISSP\AffiliationUTPhys
\affiliation{Institute for Molecular Science, Okazaki, Aichi 444-8585, Japan}
\author{Tomoyuki~Tsuyama}\AffiliationISSP
\author{Suguru~Ito}\AffiliationISSP\AffiliationUTPhys
\author{Kou~Takubo}\AffiliationISSP
\author{Iwao~Matsuda}\AffiliationISSP\AffiliationUTPhys
\author{Niko~Pontius}\AffiliationHZB
\author{Christian~Sch\"{u}\ss ler-Langeheine}\AffiliationHZB
\author{Makoto~Minohara}\AffiliationPF
\affiliation{Research Institute for Advanced Electronics and Photonics,
National Institute of Advanced Industrial Science and Technology (AIST),
Tsukuba, Ibaraki 305-8568, Japan}
\author{Hiroshi~Kumigashira}\AffiliationPF
\affiliation{Institute of Multidisciplinary Research for Advanced Materials (IMRAM), Tohoku University, Sendai 980?8577, Japan}
\author{Yuichi~Yamasaki}\AffiliationPF
\affiliation{National Institute for Materials Science (NIMS), Tsukuba 305-0047, Japan}
\affiliation{PRESTO, Japan Science and Technology Agency (JST), Saitama 332-0012, Japan}
\affiliation{RIKEN Center for Emergent Matter Science (CEMS), Wako 351-0198, Japan}
\author{Hironori~Nakao}\AffiliationPF
\author{Youichi~Murakami}\AffiliationPF
\author{Takayoshi~Katase}\AffiliationTitech
\author{Toshio~Kamiya}\AffiliationTitech
\author{Hiroki~Wadati}
\AffiliationISSP\AffiliationUTPhys
\affiliation{Department of Material Science, University of Hyogo, Kamigori-cho, Hyogo 678-1297, Japan}
\affiliation{Institute of Laser Engineering, Osaka University, Suita, Osaka 565-0871, Japan}
\date{\today}
\begin{abstract}
The relationship between the magnetic interaction and photoinduced dynamics in antiferromagnetic perovskites is investigated in this study.
In \LSFO thin films, commensurate spin ordering is accompanied by charge disproportionation, whereas \SFO thin films show incommensurate helical antiferromagnetic spin ordering due to increased ferromagnetic coupling compared to \LSFO.
To understand the photoinduced spin dynamics in these materials, we investigate the spin ordering through time-resolved resonant soft X-ray scattering.
In \LSFO, ultrafast quenching of the magnetic ordering within 130 fs through a nonthermal process is observed, triggered by charge transfer between the Fe atoms. 
We compare this to the photoinduced dynamics of the helical magnetic ordering of \SFO.
We find that the change in the magnetic coupling through optically induced charge transfer can offer an even more efficient channel for spin-order manipulation.
\end{abstract}
\maketitle

\section{Introduction}

\begin{figure}
  \begin{center} 
    \includegraphics[clip,width=7cm]{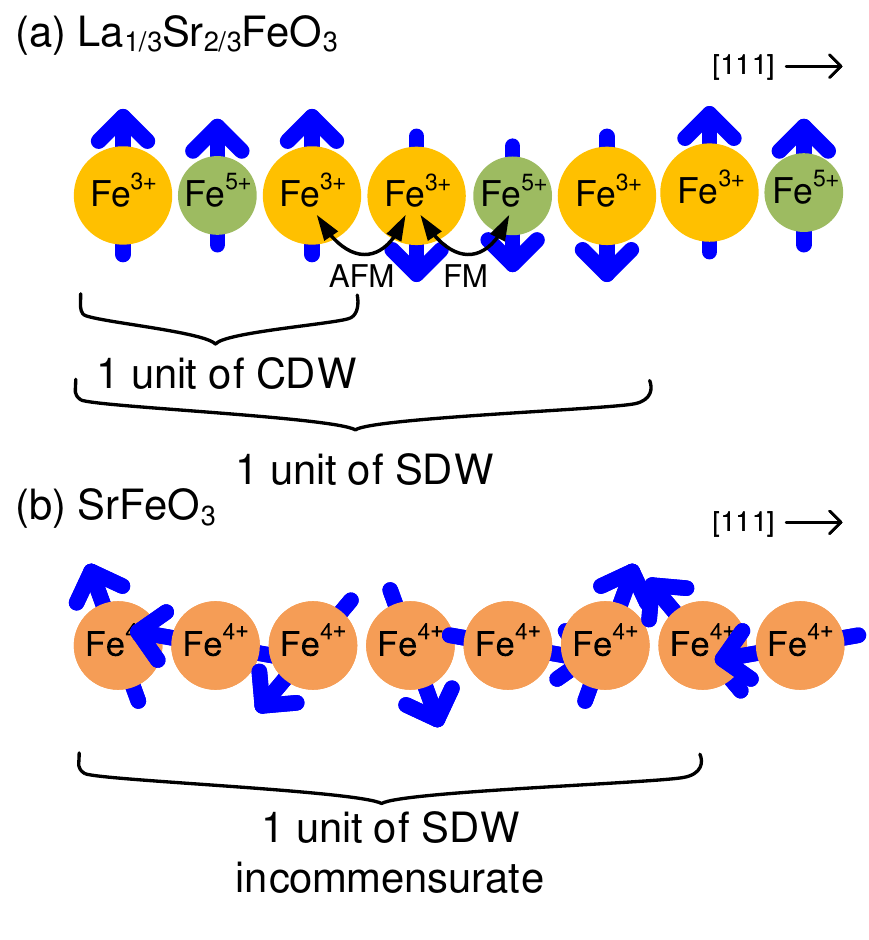}
    \caption{Schematic of the spin and charge ordering in (a) \LSFO and (b)\SFO
    (only Fe ions along the 111 direction are depicted).}
    \label{fig:order} 
  \end{center}
\end{figure}

In the last century, the challenges in material science have mainly included the observation and manipulation of charges in materials, for application in semiconductor electronics. 
In the past decades, the utilization of spins, i.e., spintronics, has attracted considerable interest.
The faster and energy-efficient manipulation of spins is one of the main issues in spintronics. 
Optical control of spins with ultrashort laser pulses is an important research topic for the ultrafast control of magnetization~\cite{Kirilyuk2010}.
The first reported ultrafast control involved photoinduced demagnetization in ferromagnetic Ni within 1 ps~\cite{Beaurepaire1996}. Subsequently, the photoinduced charge and spin dynamics have been studied extensively through diffraction and spectroscopy in ferro-, antiferro-, and ferrimagnetic materials~\cite{Stamm2007,Stanciu2007,Koopmans2009a,Radu2011,Johnson2012,Lee2012,Caviglia2013,Beaud2014,Forst2014,Forst2015,Tsuyama2016}.

Antiferromagnets are expected to be the key material for the photocontrol of magnetic orderings.
In the case of ferromagnetic materials, the magnetic moments need to be transferred from the spin to the other degrees-of-freedom, such as the lattice, when the magnetic moments are changed.
However, due to the quenched total magnetic moments of antiferromagnets, angular momentum transfer between the spin and other systems is not necessary in antiferromagnets, enabling ultrafast magnetization changes.
This was clarified by comparing the ferro- and antiferromagnetic phases of Dy~\cite{Thielemann-Kuhn2017a}.

Several perovskite oxides with antiferromagnetic orderings have been discovered, whose properties can be controlled through elemental substitution and doping because of the strong interactions between the electrons, lattice, and spin degree-of- freedom~\cite{Imada1998}.
Taking advantage of this feature, it is expected that perovskite oxide materials suitable for the fast optical control of magnetism can be discovered.

The equilibrium magnetic properties of \ce{La_{1-$x$}Sr_$x$FeO3} perovskites can be controlled based on the La/Sr composition.
\LSFO thin films exhibit a charge disproportionation (CD) of 3Fe$^{3.67+}$ $\rightarrow$ Fe$^{5+}$ + 2Fe$^{3+}$ accompanying antiferromagnetic ordering below $T_\mathrm{N}=T_\mathrm{CD}\approx$ 190~K \cite{Wadati2005,Sichel-Tissot2013}.
The high valence state of \ce{Fe^{5+}} is realized as $\mathrm{Fe}^{3+}\underline{L}^2$, where $\underline{L}$ denotes an oxygen 2\textit{p} hole~\cite{Abbate1992}.
This charge disproportionation phase results in a charge density wave (CDW) with threefold periodicity as that of the crystal lattice, and a spin density wave (SDW) of sixfold periodicity along the 111 directions, as revealed by a neutron diffraction study (see Fig.~\ref{fig:order} (a))\cite{Imada1998}. 
\SFO includes the \ce{Fe^{4+}} ion and shows a helimagnetic phase with incommensurate periodicity,
which occurs because of the competition between the nearest neighbor ferromagnetic and the next-nearest neighbor antiferromagnetic interaction of the conducting 3\textit{d} electrons, as indicated by neutron scattering results~\cite{Takeda1972}.
$T_\mathrm{N}$ was reported to be 134~K ~\cite{Takeda1972} (bulk), $\approx105$~K~\cite{Chakraverty2013}, $\approx115$~K~\cite{Rogge2019} (thin films).
A multiple $Q$ helimagnetic phase in \SFO was argued~\cite{Chakraverty2013,Ishiwata2011}, which could possibly host skyrmion crystals~\cite{Ishiwata2020}.

Resonant soft X-ray scattering (RSXS)~\cite{Fink2013} is a powerful tool for revealing the ordered structures in solids, such as the magnetic, charge, and orbital ordering~\cite{Schuessler-Langeheine2001,Zhou2011,Partzsch2011,Wadati2012,Fink2013,Matsuda2015a,Yamamoto2018}.
RSXS can be performed with the core level absorption process, and the interaction between X-rays and a specific element can be enhanced. 
Hence, RSXS can be applied to thin films with small sample volumes.
X-ray polarization is sensitive to the orbitals of the valence electrons and RSXS can detect orbital and spin ordering, which are coupled through spin-orbital interaction.
The magnetic ordering of irons in \LSFO~\cite{Okamoto2010,Yamamoto2018} and \SFO~\cite{Chakraverty2013} have been observed through RSXS at the Fe L${}_{2,3}$ absorption edge (2\textit{p}$\rightarrow$ 3\textit{d}, $\approx709$~eV).

In this study, we report the photoinduced magnetic dynamics of \LSFO and \SFO thin films determined through time-resolved RSXS measurements.
We ascertain the ultrafast melting of the \LSFO magnetic ordering through charge transfer between Fe ions, which can be attributed to the strong coupling between the charge and spin in the system.
Based on the comparison between \LSFO and \SFO, the effect of the electronic properties on the dynamics of the photoinduced quenching of the magnetic orderings is examined.

\section{Experimental}

\begin{figure}
  \begin{center}
    \includegraphics[clip,width=5cm]{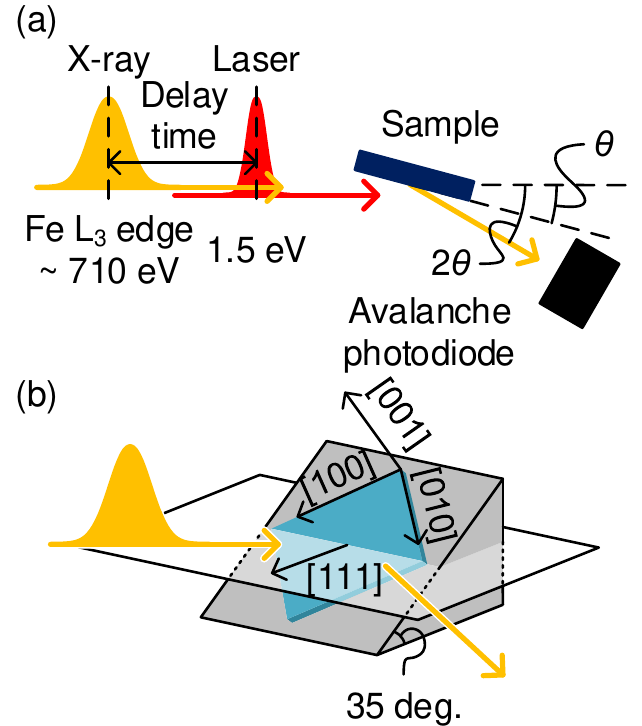}
    \caption{(a) Experimental setup for time-resolved RSXS using the pump-probe method and
    (b) geometry for the \SFO RSXS measurements.}
   \label{fig:setup}
    \end{center} 
\end{figure}

\begin{figure*}
  \begin{center}
    \includegraphics[clip,width=15cm]{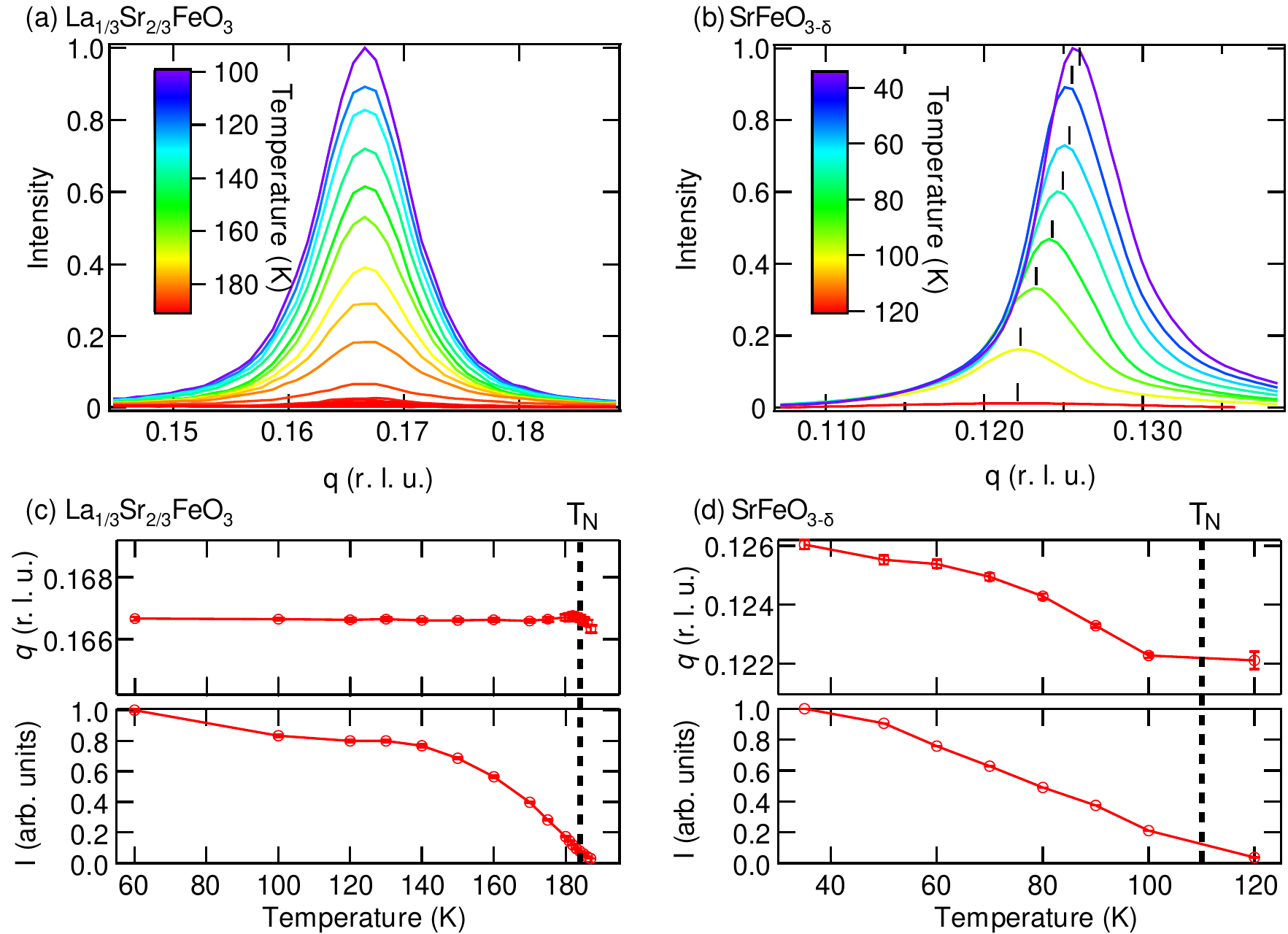}
    \caption{Temperature dependence of the diffraction peak of (a) \LSFO and (b)\SFO; 
    (c) \LSFO peak position $q$ and intensity $I$, and (d) \SFO plotted as a function of the temperature.}
    \label{fig:diff}
  \end{center}
\end{figure*}

\LSFO thin films with $\approx$ 40-nm thicknesses were grown epitaxially on \ce{SrTiO3} (111) substrates using the pulsed laser deposition method.
The details can be found in Ref.~\cite{Minohara2016}.
By oxidizing \ce{SrFeO_{2.5}}, \SFO thin films with $\approx27$~nm thicknesses were fabricated on \ce{SrTiO3}(100) substrates using the pulsed laser deposition method.
\SFO was obtained by annealing SrFeO${}_{2.5}$ in an ozone atmosphere at 300~${}^\circ\textrm{C}$ for 6 h with the sample exposed to UV light in a UV/O3 DRY CLEANER UV-1 (SAMCO Inc).

In order to elucidate the photoinduced dynamics of the antiferromagnetic ordering in \LSFO and \SFO, we performed time-resolved RSXS measurements using the pump-probe method at the slicing facility UE56/1-ZPM Helmholtz-Zentrum, Berlin~\cite{Holldack2014}
The experimental setup is shown in Fig.~\ref{fig:setup}.
A Ti:Sapphire laser ($\lambda=800~\mathrm{nm}$, 1.5~eV) was employed for photoexcitation, with a pulse duration of 50~fs.
The laser and X-ray pulse frequencies were 3 and 6~kHz, respectively, and signals with and without pump laser excitation were obtained alternately.
The signals without excitation were used for normalizing the pumped signals.
The scattered X-rays were detected using an avalanche photo diode.
\LSFO was investigated using 100-fs X-ray pulses generated through the laser slicing technique~\cite{Holldack2014}; the total time resolution was $\approx 130~\mathrm{fs}$.
\SFO was mounted on a wedge-shaped jig at an angle of 55 deg. in order to orient the [111] direction in the scattering plane, and the temperature was set to 35~K.
We set the X-ray photon energy and polarization to the Fe L${}_3$ edge ($\approx 710$ eV) and horizontal, respectively, for obtaining the magnetic signals. 
In the \SFO case, the time resolution was approximately 50~ps, corresponding to the pulse width of the synchrotron X-ray radiation.
Static \LSFO measurements were performed with a diffractometer at the Soft X-ray beamline BL-16A, Photon Factory, KEK, Japan~\cite{Nakao2014}.
The experimental geometry and temperature were equivalent to the time-resolved measurement, and a silicon drift detector was used.

\section{Results and Discussion}

Figure~\ref{fig:diff} depicts the $Q=(q,q,q)$ antiferromagnetic ordering peaks of (a) \LSFO and (b) \SFO.
The diffraction peaks were scanned along the [111] direction.
The temperature dependence of $q$ and the peak intensities of \LSFO and \SFO are displayed in Figs.~\ref{fig:diff}(c,d) respectively.
For the \LSFO thin films, the peak position is fixed at $(1/6,1/6,1/6)$ in the entire temperature range below $T_N$, as shown in Fig~\ref{fig:diff}(a), and $T_\mathrm{N}$ is estimated to be 182~K.
The \SFO diffraction peak appears below $T_\mathrm{N}\approx 110~\mathrm{K}$, and the peak position shifts according to the temperature.
$T_N\approx110~\textrm{K}$ is also in the same range as those of previous reports~\cite{Chakraverty2013,Rogge2019}.
$T_\mathrm{N}$ depends on the oxygen vacancy $\delta$, and decreases from 134 K for $\delta=0$ to 80~K for $\delta=0.16$~\cite{MacChesney1965}, which implies a small $\delta$ for our sample.
$q$ of the \SFO helimagnetic ordering was reported to be 0.128-0.112 (\ce{SrFeO_{2.87}})~\cite{Reehuis2012,Takeda1972}, 0.13 (\SFO) or 0.125 (\ce{Sr_{0.99}Co_{0.01}FeO3})~\cite{Chakraverty2013}; our observed $q=0.126$ is in a similar range.
In the heating and cooling cycle, thermal hysteresis was observed for the peak intensity and peak positions.
According to previous resistivity measurement reports~\cite{Chakraverty2013,Rogge2019}, thermal hysteresis was observed at $T=46-71~K$, and our observed thermal hysteresis reflects this phenomenon.

\begin{figure}
  \begin{center}
    \includegraphics[clip,width=8cm]{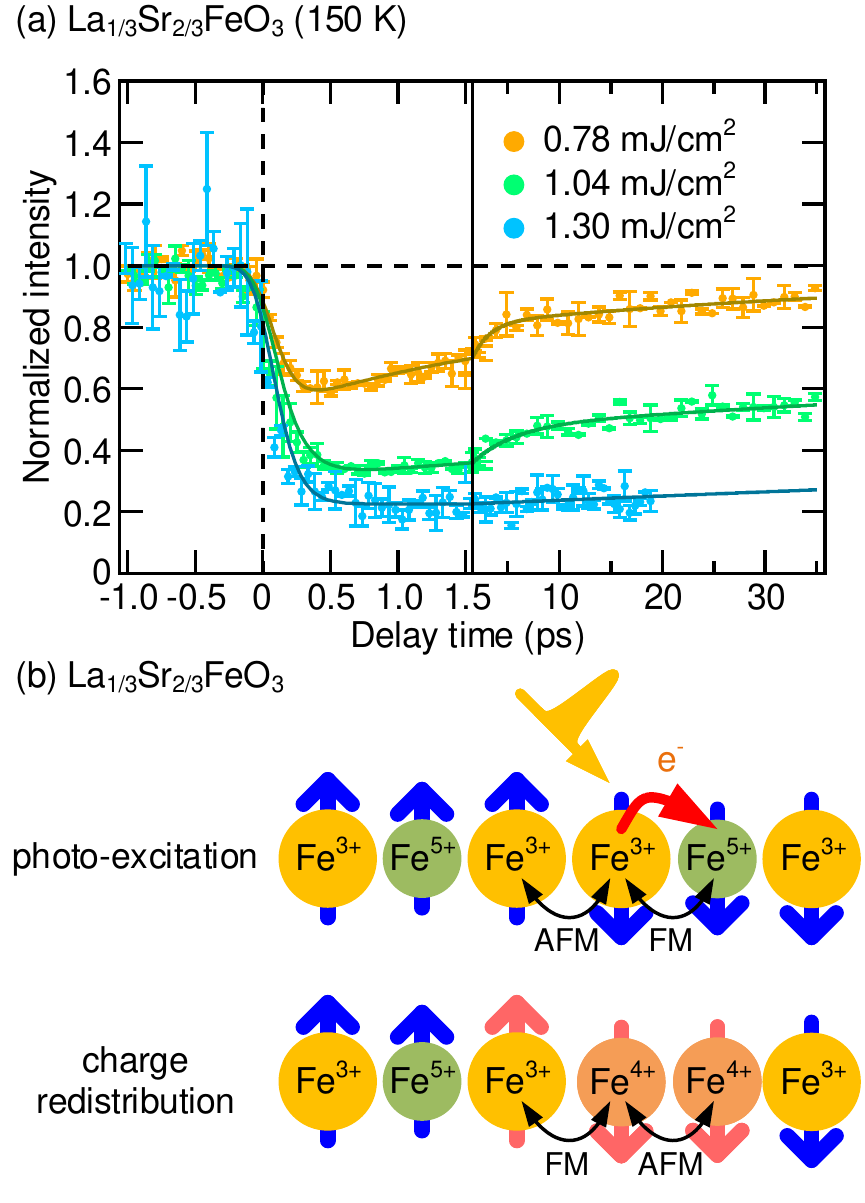}
    \caption{(a) Time evolution of the magnetic peak intensity of \LSFO at $\tau_\mathrm{resolution}=130~\mathrm{fs}$ for various pump laser fluence obtained in the laser slicing mode (the solid lines are fitted to the data using Eq.~\eqref{eq:LSFOfit}) and
    (b) illustration of the photoinduced transient states of \LSFO. 
    }
    \label{fig:LSFO}
  \end{center}
\end{figure}

We first discuss the \LSFO dynamics. 
The time evolution of the magnetic peak intensity after laser pumping, revealed in the laser slicing mode, is shown in Fig.~\ref{fig:LSFO}(a).
Rapid reduction in the diffraction intensity is observed within a time resolution of 130~fs.
The decreased diffraction intensity recovers 30 ps after laser excitation in the lower fluence regime; however, the recovery is slower at higher fluence. 
In order to discuss the results quantitatively, we fitted them using the function depicted by Eq.~\eqref{eq:LSFOfit} for the decay and recovery processes.
The fitting function is as follows:
\begin{multline}
  f(t)=1+\left[\left(R_\mathrm{fast}+R_\mathrm{slow}\right)\exp\left(\frac{-t}{\tau_\mathrm{decay}}\right) \right. \\
  \left. -R_\mathrm{fast}\exp\left(\frac{-t}{\tau_\mathrm{fast}}\right)-R_\mathrm{slow}\exp\left(\frac{-t}{\tau_\mathrm{slow}}\right)\right]H(t) \\
  \ast\exp\left(-\frac{\ln(2)}{4}\frac{t^2}{\tau_\mathrm{resolution}^2}\right), \label{eq:LSFOfit}
\end{multline}
where $\tau_\mathrm{decay}$, $\tau_\mathrm{slow}$, and $\tau_\mathrm{fast}$ are the corresponding time constants, $\ast$ denotes the convolution, and $H(t)$ is the Heaviside step function.
To consider time resolution, the fitting function was convoluted by a Gaussian with a full-width-at-half-maximum of $\tau_\mathrm{resolution} = 130~\mathrm{fs}$.
The parameters extracted from the experimental delay scans are summarized in Table~\ref{LSFOfit}.
The $\tau_\mathrm{decay}$ time scale was estimated to be $0.1\pm0.05$~ps.

For the nonthermal \LSFO process with an ultrashort time scale of $\tau_\mathrm{decay}\lesssim0.1~\mathrm{ps}$, we consider that the ultrafast magnetization dynamics occurs because of the ultrafast photo-melting of the charge ordering, which destroys the magnetic ordering due to the strong coupling between the charge and spin.
Demagnetization due to the photo-melting of the charge ordering has also been observed, for example, in NdNiO$_3$ thin films~\cite{Caviglia2013}.
Figure \ref{fig:LSFO}(b) illustrates the origins of the ultrafast melting of the magnetic ordering.
Due to the strong onsite Hund's rule coupling, excitation with linearly polarized light at 1.5 eV can result only in transitions between the ferromagnetically coupled sites~\cite{Ishikawa1998}.
Optical conductivity measurement has demonstrated that the optical gap of 0.15~eV has a strong interatomic $d-d$ transition characteristic~\cite{Ishikawa1998}.
Therefore, the excited electron would be transferred only in the ferromagnetically coupled Fe$^{3+}$ - Fe$^{5+}$ - Fe$^{3+}$ sites, and the observation in this study suggests that the excitation induces charge transfer in a unit of the charge-ordering site of Fe$^{3+}$ - Fe$^{5+}$ - Fe$^{3+}$ into Fe$^{3+}$ - Fe$^{4+}$ - Fe$^{4+}$, as shown in Fig.~\ref{fig:LSFO}(b).
It has been suggested that this charge transfer is a metastable state accompanying Fe-O bond-length change, in a previous report on the photoinduced transient states of \LSFO~\cite{Zhu2018}.
Transfer in the charge ordering induces changes in the magnetic interactions.
Fe$^{3+}$ - Fe$^{4+}$ ions have ferromagnetic interactions, whereas the Fe$^{4+}$ - Fe$^{4+}$ sites have antiferromagnetic ones, as shown in Fig.~\ref{fig:LSFO}(b).

\begin{figure}
  \begin{center}
    \includegraphics[clip,width=8cm]{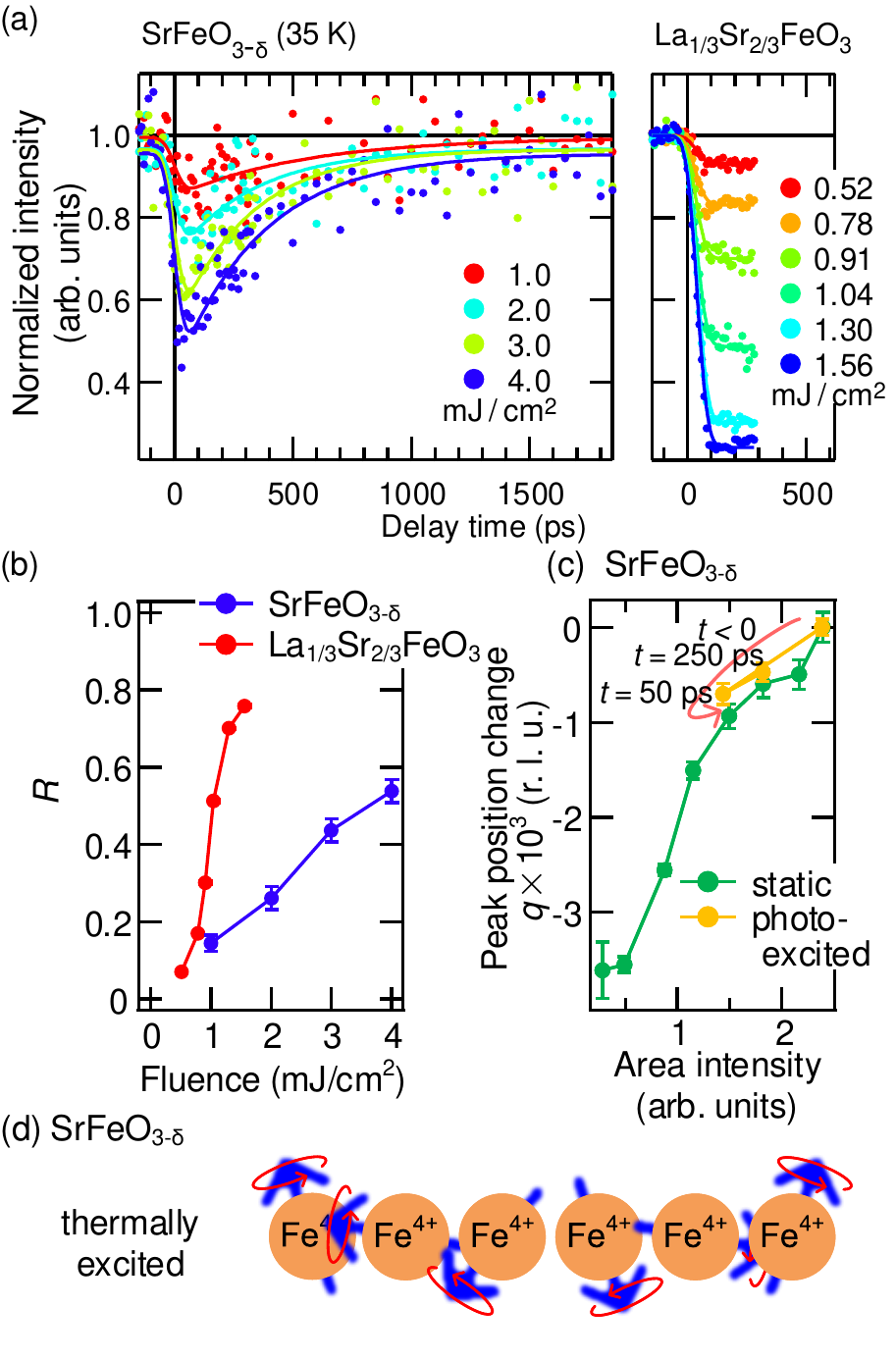}
    \caption{
(a) Pump-probe delay scans of \SFO (right panel) and \LSFO (left panel) $\tau_\mathrm{resolution}=50~\mathrm{ps}$, (b) degree of photoinduced quenching of the diffraction peak intensity as a function of the laser fluence, (c) magnetic peak position of the photoinduced transient states and static states of \SFO as a function of the peak area intensity, and (d) Illustration of the photoinduced transient states of \SFO. 
    }
  \label{fig:SFO}
  \end{center}
\end{figure}

Furthermore, time-resolved RSXS measurements were performed for 35 and 80~K \SFO thin films as shown in Fig.~\ref{fig:SFO}(a) with a time resolution of $\tau_\mathrm{resolution}\approx 50~\mathrm{ps}$; 
the experimental results for \LSFO with a similar time resolution are also shown in the right panel for comparison.
Due to $\tau_\mathrm{resolution}\approx50~\mathrm{ps}$, the fast decay process cannot be resolved in Fig.~\ref{fig:SFO}(a), and we focus on the changes in the magnetic ordering peak intensity. 
We fitted using the following single exponential function
\begin{multline}
 f(t)=\left[1-R_\mathrm{slow}\exp\left(\frac{-t}{\tau_\mathrm{slow}}\right)H(t)\right] \ast\exp\left(-\frac{\ln(2)}{4}\frac{t^2}{\tau_\mathrm{resolution}^2}\right), \label{eq:LSFOSFOfit}
\end{multline}
which is the limit of Eq.~\eqref{eq:LSFOfit} as $R_\mathrm{fast}$, $\tau_\mathrm{decay}$, and $\tau_\mathrm{fast} \rightarrow 0$.

\begin{table}
\begin{center}
\caption{Fitting parameters for the delay scans of the \LSFO RSXS intensities in Fig.~\ref{fig:LSFO}(a) observed in the slicing mode.
$\tau_\mathrm{resolution} =$ 130 fs.
$\tau_\mathrm{decay}$ is less than time resolution $\tau_\mathrm{resolution}$.}
\begin{tabular}{c|ccc}
\hline \hline
Fluence [mJ/cm$^2$] & $R$ & $\tau_{fast}$~[ps] & $\tau_{slow}$~[ps] \\
\hline
0.78 & 0.51 & 1.7 & 60 \\ 
1.04 & 0.72 & 3.5 & 210 \\
1.30 & 0.78 & - & 550 \\
\hline \hline
\end{tabular}
\label{LSFOfit}
\end{center}
\end{table}

The degrees of quenching $R_\mathrm{slow}$, defined by Eq.~\eqref{eq:LSFOSFOfit}, of the \LSFO and \SFO magnetic ordering are plotted as a function of the laser fluence in Fig.~\ref{fig:SFO}(b).
For \LSFO, the quenching of the peak intensity reaches $\approx~70~\%$ at a fluence of approximately $1.5~\textrm{mJ/cm}^2$; however, the \SFO diffraction intensity decreases by less than 60~\% of the initial intensity below a fluence of $4~\rm{mJ/cm}^2$.
This suggests that the antiferromagnetic ordering of \LSFO is quenched with less excitation energy compared to \SFO.
The peak positions of the photoinduced transient states and the heating process are plotted as a function of the peak area intensity in Fig.~\ref{fig:SFO}(c).
Both curves are similar, indicating that the observed photoinduced quenching of the \SFO helimagnetic ordering is triggered by the same mechanism as the temperature-induced change.

\SFO is a metallic material and 1.5~eV laser excitation corresponds to the interband transition~\cite{Fujioka2012}.
The photoinduced quenching of the helimagnetic ordering in \SFO can be interpreted as the result of the increase in the spin and electron temperatures, as depicted in Fig.~\ref{fig:SFO}(d).
On the other hand, \LSFO is insulating and its magnetic orderings disappear because of the charge transfer between Fe ions induced locally by photo excitation, as discussed above. 
The photoinduced dynamics of the antiferromagnetic orderings reflect the versatile electronic feature of perovskite oxides.

\section{Conclusion}
In this study, we examined the photoinduced dynamics of the antiferromagnetic orderings in two perovskites, \LSFO and \SFO, through  time-resolved RSXS.
The magnetic ordering in \LSFO thin films was quenched within 130~fs.
This ultrafast dynamics can be explained based on the photoinduced charge transfer between Fe ions, which induces a change in the strong magnetic interactions. The process is nonthermal. 
The spin moment can be cancelled with the nearest neighbors in antiferromagnetic \LSFO through this ultrafast process, leading to ultrafast photoinduced change in the magnetic ordering.
Compared to helimagnetic \SFO, the magnetic ordering of \LSFO was destroyed at lower fluence.
Spin manipulation in helical systems has been found to be more energy efficient than in ferromagnets.
An even more energy efficient process can be realized in \LSFO.
Ultrafast changes in the magnetic ordering of antiferromagnetic perovskite thin films were discovered, and the photoinduced dynamics was shown to be controlled by tuning the electronic feature through perovskite doping. 
The photoinduced dynamics of the spin order in antiferromagnetic perovskite thin films is important for integrating functional oxides into spintronic device architectures.

\begin{acknowledgments}
We thank Yasutuki~Hirata for the productive discussions and HZB for the allocation of the synchrotron radiation beamtime, Karsten~Holldack and Rolf~Mitzner for experimental support.
This work was performed under the approval of the Photon Factory Program Advisory Committee (Proposal Nos. 2016PF-BL-19B, 2015G556, 2015S2-007, 2013G058, 2013G661).
This work was partially supported by the Japan Society for the Promotion of Science (JSPS) KAKENHI Grant Nos. 19H05824, 19H01816, 19K23430, and 17K14334, and the MEXT Quantum Leap Flagship Program (MEXT Q-LEAP) Grant No. JPMXS0118068681.
K.~Y. acknowledges the support from the ALPS program of the University of Tokyo.
\end{acknowledgments}

\bibliography{/home/yamako/Dropbox/library}

\begin{thebibliography}{39}%
\makeatletter
\providecommand \@ifxundefined [1]{%
 \@ifx{#1\undefined}
}%
\providecommand \@ifnum [1]{%
 \ifnum #1\expandafter \@firstoftwo
 \else \expandafter \@secondoftwo
 \fi
}%
\providecommand \@ifx [1]{%
 \ifx #1\expandafter \@firstoftwo
 \else \expandafter \@secondoftwo
 \fi
}%
\providecommand \natexlab [1]{#1}%
\providecommand \enquote  [1]{``#1''}%
\providecommand \bibnamefont  [1]{#1}%
\providecommand \bibfnamefont [1]{#1}%
\providecommand \citenamefont [1]{#1}%
\providecommand \href@noop [0]{\@secondoftwo}%
\providecommand \href [0]{\begingroup \@sanitize@url \@href}%
\providecommand \@href[1]{\@@startlink{#1}\@@href}%
\providecommand \@@href[1]{\endgroup#1\@@endlink}%
\providecommand \@sanitize@url [0]{\catcode `\\12\catcode `\$12\catcode
  `\&12\catcode `\#12\catcode `\^12\catcode `\_12\catcode `\%12\relax}%
\providecommand \@@startlink[1]{}%
\providecommand \@@endlink[0]{}%
\providecommand \url  [0]{\begingroup\@sanitize@url \@url }%
\providecommand \@url [1]{\endgroup\@href {#1}{\urlprefix }}%
\providecommand \urlprefix  [0]{URL }%
\providecommand \Eprint [0]{\href }%
\providecommand \doibase [0]{http://dx.doi.org/}%
\providecommand \selectlanguage [0]{\@gobble}%
\providecommand \bibinfo  [0]{\@secondoftwo}%
\providecommand \bibfield  [0]{\@secondoftwo}%
\providecommand \translation [1]{[#1]}%
\providecommand \BibitemOpen [0]{}%
\providecommand \bibitemStop [0]{}%
\providecommand \bibitemNoStop [0]{.\EOS\space}%
\providecommand \EOS [0]{\spacefactor3000\relax}%
\providecommand \BibitemShut  [1]{\csname bibitem#1\endcsname}%
\let\auto@bib@innerbib\@empty
\bibitem [{\citenamefont {Kirilyuk}\ \emph {et~al.}(2010)\citenamefont
  {Kirilyuk}, \citenamefont {Kimel},\ and\ \citenamefont
  {Rasing}}]{Kirilyuk2010}%
  \BibitemOpen
  \bibfield  {author} {\bibinfo {author} {\bibfnamefont {A.}~\bibnamefont
  {Kirilyuk}}, \bibinfo {author} {\bibfnamefont {A.~V.}\ \bibnamefont {Kimel}},
  \ and\ \bibinfo {author} {\bibfnamefont {T.}~\bibnamefont {Rasing}},\ }\href
  {\doibase 10.1103/RevModPhys.82.2731} {\bibfield  {journal} {\bibinfo
  {journal} {Rev. Mod. Phys.}\ }\textbf {\bibinfo {volume} {82}},\ \bibinfo
  {pages} {2731} (\bibinfo {year} {2010})}\BibitemShut {NoStop}%
\bibitem [{\citenamefont {Beaurepaire}\ \emph {et~al.}(1996)\citenamefont
  {Beaurepaire}, \citenamefont {Merle}, \citenamefont {Daunois},\ and\
  \citenamefont {Bigot}}]{Beaurepaire1996}%
  \BibitemOpen
  \bibfield  {author} {\bibinfo {author} {\bibfnamefont {E.}~\bibnamefont
  {Beaurepaire}}, \bibinfo {author} {\bibfnamefont {J.-C.}\ \bibnamefont
  {Merle}}, \bibinfo {author} {\bibfnamefont {A.}~\bibnamefont {Daunois}}, \
  and\ \bibinfo {author} {\bibfnamefont {J.-Y.}\ \bibnamefont {Bigot}},\ }\href
  {\doibase 10.1103/PhysRevLett.76.4250} {\bibfield  {journal} {\bibinfo
  {journal} {Phys. Rev. Lett.}\ }\textbf {\bibinfo {volume} {76}},\ \bibinfo
  {pages} {4250} (\bibinfo {year} {1996})}\BibitemShut {NoStop}%
\bibitem [{\citenamefont {Stamm}\ \emph {et~al.}(2007)\citenamefont {Stamm},
  \citenamefont {Kachel}, \citenamefont {Pontius}, \citenamefont {Mitzner},
  \citenamefont {Quast}, \citenamefont {Holldack}, \citenamefont {Khan},
  \citenamefont {Lupulescu}, \citenamefont {Aziz}, \citenamefont {Wietstruk},
  \citenamefont {D{\"{u}}rr},\ and\ \citenamefont {Eberhardt}}]{Stamm2007}%
  \BibitemOpen
  \bibfield  {author} {\bibinfo {author} {\bibfnamefont {C.}~\bibnamefont
  {Stamm}}, \bibinfo {author} {\bibfnamefont {T.}~\bibnamefont {Kachel}},
  \bibinfo {author} {\bibfnamefont {N.}~\bibnamefont {Pontius}}, \bibinfo
  {author} {\bibfnamefont {R.}~\bibnamefont {Mitzner}}, \bibinfo {author}
  {\bibfnamefont {T.}~\bibnamefont {Quast}}, \bibinfo {author} {\bibfnamefont
  {K.}~\bibnamefont {Holldack}}, \bibinfo {author} {\bibfnamefont
  {S.}~\bibnamefont {Khan}}, \bibinfo {author} {\bibfnamefont {C.}~\bibnamefont
  {Lupulescu}}, \bibinfo {author} {\bibfnamefont {E.~F.}\ \bibnamefont {Aziz}},
  \bibinfo {author} {\bibfnamefont {M.}~\bibnamefont {Wietstruk}}, \bibinfo
  {author} {\bibfnamefont {H.~A.}\ \bibnamefont {D{\"{u}}rr}}, \ and\ \bibinfo
  {author} {\bibfnamefont {W.}~\bibnamefont {Eberhardt}},\ }\href {\doibase
  10.1038/nmat1985} {\bibfield  {journal} {\bibinfo  {journal} {Nat. Mater.}\
  }\textbf {\bibinfo {volume} {6}},\ \bibinfo {pages} {740} (\bibinfo {year}
  {2007})}\BibitemShut {NoStop}%
\bibitem [{\citenamefont {Stanciu}\ \emph {et~al.}(2007)\citenamefont
  {Stanciu}, \citenamefont {Hansteen}, \citenamefont {Kimel}, \citenamefont
  {Kirilyuk}, \citenamefont {Tsukamoto}, \citenamefont {Itoh},\ and\
  \citenamefont {Rasing}}]{Stanciu2007}%
  \BibitemOpen
  \bibfield  {author} {\bibinfo {author} {\bibfnamefont {C.~D.}\ \bibnamefont
  {Stanciu}}, \bibinfo {author} {\bibfnamefont {F.}~\bibnamefont {Hansteen}},
  \bibinfo {author} {\bibfnamefont {A.~V.}\ \bibnamefont {Kimel}}, \bibinfo
  {author} {\bibfnamefont {A.}~\bibnamefont {Kirilyuk}}, \bibinfo {author}
  {\bibfnamefont {A.}~\bibnamefont {Tsukamoto}}, \bibinfo {author}
  {\bibfnamefont {A.}~\bibnamefont {Itoh}}, \ and\ \bibinfo {author}
  {\bibfnamefont {T.}~\bibnamefont {Rasing}},\ }\href {\doibase
  10.1103/PhysRevLett.99.047601} {\bibfield  {journal} {\bibinfo  {journal}
  {Phys. Rev. Lett.}\ }\textbf {\bibinfo {volume} {99}},\ \bibinfo {pages}
  {047601} (\bibinfo {year} {2007})}\BibitemShut {NoStop}%
\bibitem [{\citenamefont {Koopmans}\ \emph {et~al.}(2010)\citenamefont
  {Koopmans}, \citenamefont {Malinowski}, \citenamefont {{Dalla Longa}},
  \citenamefont {Steiauf}, \citenamefont {F{\"{a}}hnle}, \citenamefont {Roth},
  \citenamefont {Cinchetti},\ and\ \citenamefont
  {Aeschlimann}}]{Koopmans2009a}%
  \BibitemOpen
  \bibfield  {author} {\bibinfo {author} {\bibfnamefont {B.}~\bibnamefont
  {Koopmans}}, \bibinfo {author} {\bibfnamefont {G.}~\bibnamefont
  {Malinowski}}, \bibinfo {author} {\bibfnamefont {F.}~\bibnamefont {{Dalla
  Longa}}}, \bibinfo {author} {\bibfnamefont {D.}~\bibnamefont {Steiauf}},
  \bibinfo {author} {\bibfnamefont {M.}~\bibnamefont {F{\"{a}}hnle}}, \bibinfo
  {author} {\bibfnamefont {T.}~\bibnamefont {Roth}}, \bibinfo {author}
  {\bibfnamefont {M.}~\bibnamefont {Cinchetti}}, \ and\ \bibinfo {author}
  {\bibfnamefont {M.}~\bibnamefont {Aeschlimann}},\ }\href {\doibase
  10.1038/nmat2593} {\bibfield  {journal} {\bibinfo  {journal} {Nat. Mater.}\
  }\textbf {\bibinfo {volume} {9}},\ \bibinfo {pages} {259} (\bibinfo {year}
  {2010})}\BibitemShut {NoStop}%
\bibitem [{\citenamefont {Radu}\ \emph {et~al.}(2011)\citenamefont {Radu},
  \citenamefont {Vahaplar}, \citenamefont {Stamm}, \citenamefont {Kachel},
  \citenamefont {Pontius}, \citenamefont {D{\"{u}}rr}, \citenamefont {Ostler},
  \citenamefont {Barker}, \citenamefont {Evans}, \citenamefont {Chantrell},
  \citenamefont {Tsukamoto}, \citenamefont {Itoh}, \citenamefont {Kirilyuk},
  \citenamefont {Rasing},\ and\ \citenamefont {Kimel}}]{Radu2011}%
  \BibitemOpen
  \bibfield  {author} {\bibinfo {author} {\bibfnamefont {I.}~\bibnamefont
  {Radu}}, \bibinfo {author} {\bibfnamefont {K.}~\bibnamefont {Vahaplar}},
  \bibinfo {author} {\bibfnamefont {C.}~\bibnamefont {Stamm}}, \bibinfo
  {author} {\bibfnamefont {T.}~\bibnamefont {Kachel}}, \bibinfo {author}
  {\bibfnamefont {N.}~\bibnamefont {Pontius}}, \bibinfo {author} {\bibfnamefont
  {H.~A.}\ \bibnamefont {D{\"{u}}rr}}, \bibinfo {author} {\bibfnamefont
  {T.~A.}\ \bibnamefont {Ostler}}, \bibinfo {author} {\bibfnamefont
  {J.}~\bibnamefont {Barker}}, \bibinfo {author} {\bibfnamefont {R.~F.~L.}\
  \bibnamefont {Evans}}, \bibinfo {author} {\bibfnamefont {R.~W.}\ \bibnamefont
  {Chantrell}}, \bibinfo {author} {\bibfnamefont {A.}~\bibnamefont
  {Tsukamoto}}, \bibinfo {author} {\bibfnamefont {A.}~\bibnamefont {Itoh}},
  \bibinfo {author} {\bibfnamefont {A.}~\bibnamefont {Kirilyuk}}, \bibinfo
  {author} {\bibfnamefont {T.}~\bibnamefont {Rasing}}, \ and\ \bibinfo {author}
  {\bibfnamefont {A.~V.}\ \bibnamefont {Kimel}},\ }\href {\doibase
  10.1038/nature09901} {\bibfield  {journal} {\bibinfo  {journal} {Nature}\
  }\textbf {\bibinfo {volume} {472}},\ \bibinfo {pages} {205} (\bibinfo {year}
  {2011})}\BibitemShut {NoStop}%
\bibitem [{\citenamefont {Johnson}\ \emph {et~al.}(2012)\citenamefont
  {Johnson}, \citenamefont {de~Souza}, \citenamefont {Staub}, \citenamefont
  {Beaud}, \citenamefont {M{\"{o}}hr-Vorobeva}, \citenamefont {Ingold},
  \citenamefont {Caviezel}, \citenamefont {Scagnoli}, \citenamefont
  {Schlotter}, \citenamefont {Turner}, \citenamefont {Krupin}, \citenamefont
  {Lee}, \citenamefont {Chuang}, \citenamefont {Patthey}, \citenamefont
  {Moore}, \citenamefont {Lu}, \citenamefont {Yi}, \citenamefont {Kirchmann},
  \citenamefont {Trigo}, \citenamefont {Denes}, \citenamefont {Doering},
  \citenamefont {Hussain}, \citenamefont {Shen}, \citenamefont {Prabhakaran},\
  and\ \citenamefont {Boothroyd}}]{Johnson2012}%
  \BibitemOpen
  \bibfield  {author} {\bibinfo {author} {\bibfnamefont {S.~L.}\ \bibnamefont
  {Johnson}}, \bibinfo {author} {\bibfnamefont {R.~A.}\ \bibnamefont
  {de~Souza}}, \bibinfo {author} {\bibfnamefont {U.}~\bibnamefont {Staub}},
  \bibinfo {author} {\bibfnamefont {P.}~\bibnamefont {Beaud}}, \bibinfo
  {author} {\bibfnamefont {E.}~\bibnamefont {M{\"{o}}hr-Vorobeva}}, \bibinfo
  {author} {\bibfnamefont {G.}~\bibnamefont {Ingold}}, \bibinfo {author}
  {\bibfnamefont {A.}~\bibnamefont {Caviezel}}, \bibinfo {author}
  {\bibfnamefont {V.}~\bibnamefont {Scagnoli}}, \bibinfo {author}
  {\bibfnamefont {W.~F.}\ \bibnamefont {Schlotter}}, \bibinfo {author}
  {\bibfnamefont {J.~J.}\ \bibnamefont {Turner}}, \bibinfo {author}
  {\bibfnamefont {O.}~\bibnamefont {Krupin}}, \bibinfo {author} {\bibfnamefont
  {W.-S.}\ \bibnamefont {Lee}}, \bibinfo {author} {\bibfnamefont {Y.-D.}\
  \bibnamefont {Chuang}}, \bibinfo {author} {\bibfnamefont {L.}~\bibnamefont
  {Patthey}}, \bibinfo {author} {\bibfnamefont {R.~G.}\ \bibnamefont {Moore}},
  \bibinfo {author} {\bibfnamefont {D.}~\bibnamefont {Lu}}, \bibinfo {author}
  {\bibfnamefont {M.}~\bibnamefont {Yi}}, \bibinfo {author} {\bibfnamefont
  {P.~S.}\ \bibnamefont {Kirchmann}}, \bibinfo {author} {\bibfnamefont
  {M.}~\bibnamefont {Trigo}}, \bibinfo {author} {\bibfnamefont
  {P.}~\bibnamefont {Denes}}, \bibinfo {author} {\bibfnamefont
  {D.}~\bibnamefont {Doering}}, \bibinfo {author} {\bibfnamefont
  {Z.}~\bibnamefont {Hussain}}, \bibinfo {author} {\bibfnamefont {Z.-X.}\
  \bibnamefont {Shen}}, \bibinfo {author} {\bibfnamefont {D.}~\bibnamefont
  {Prabhakaran}}, \ and\ \bibinfo {author} {\bibfnamefont {A.~T.}\ \bibnamefont
  {Boothroyd}},\ }\href {\doibase 10.1103/PhysRevLett.108.037203} {\bibfield
  {journal} {\bibinfo  {journal} {Phys. Rev. Lett.}\ }\textbf {\bibinfo
  {volume} {108}},\ \bibinfo {pages} {037203} (\bibinfo {year}
  {2012})}\BibitemShut {NoStop}%
\bibitem [{\citenamefont {Lee}\ \emph {et~al.}(2012)\citenamefont {Lee},
  \citenamefont {Chuang}, \citenamefont {Moore}, \citenamefont {Zhu},
  \citenamefont {Patthey}, \citenamefont {Trigo}, \citenamefont {Lu},
  \citenamefont {Kirchmann}, \citenamefont {Krupin}, \citenamefont {Yi},
  \citenamefont {Langner}, \citenamefont {Huse}, \citenamefont {Robinson},
  \citenamefont {Chen}, \citenamefont {Zhou}, \citenamefont {Coslovich},
  \citenamefont {Huber}, \citenamefont {Reis}, \citenamefont {Kaindl},
  \citenamefont {Schoenlein}, \citenamefont {Doering}, \citenamefont {Denes},
  \citenamefont {Schlotter}, \citenamefont {Turner}, \citenamefont {Johnson},
  \citenamefont {F{\"{o}}rst}, \citenamefont {Sasagawa}, \citenamefont {Kung},
  \citenamefont {Sorini}, \citenamefont {Kemper}, \citenamefont {Moritz},
  \citenamefont {Devereaux}, \citenamefont {Lee}, \citenamefont {Shen},\ and\
  \citenamefont {Hussain}}]{Lee2012}%
  \BibitemOpen
  \bibfield  {author} {\bibinfo {author} {\bibfnamefont {W.}~\bibnamefont
  {Lee}}, \bibinfo {author} {\bibfnamefont {Y.}~\bibnamefont {Chuang}},
  \bibinfo {author} {\bibfnamefont {R.}~\bibnamefont {Moore}}, \bibinfo
  {author} {\bibfnamefont {Y.}~\bibnamefont {Zhu}}, \bibinfo {author}
  {\bibfnamefont {L.}~\bibnamefont {Patthey}}, \bibinfo {author} {\bibfnamefont
  {M.}~\bibnamefont {Trigo}}, \bibinfo {author} {\bibfnamefont
  {D.}~\bibnamefont {Lu}}, \bibinfo {author} {\bibfnamefont {P.}~\bibnamefont
  {Kirchmann}}, \bibinfo {author} {\bibfnamefont {O.}~\bibnamefont {Krupin}},
  \bibinfo {author} {\bibfnamefont {M.}~\bibnamefont {Yi}}, \bibinfo {author}
  {\bibfnamefont {M.}~\bibnamefont {Langner}}, \bibinfo {author} {\bibfnamefont
  {N.}~\bibnamefont {Huse}}, \bibinfo {author} {\bibfnamefont {J.}~\bibnamefont
  {Robinson}}, \bibinfo {author} {\bibfnamefont {Y.}~\bibnamefont {Chen}},
  \bibinfo {author} {\bibfnamefont {S.}~\bibnamefont {Zhou}}, \bibinfo {author}
  {\bibfnamefont {G.}~\bibnamefont {Coslovich}}, \bibinfo {author}
  {\bibfnamefont {B.}~\bibnamefont {Huber}}, \bibinfo {author} {\bibfnamefont
  {D.}~\bibnamefont {Reis}}, \bibinfo {author} {\bibfnamefont {R.}~\bibnamefont
  {Kaindl}}, \bibinfo {author} {\bibfnamefont {R.}~\bibnamefont {Schoenlein}},
  \bibinfo {author} {\bibfnamefont {D.}~\bibnamefont {Doering}}, \bibinfo
  {author} {\bibfnamefont {P.}~\bibnamefont {Denes}}, \bibinfo {author}
  {\bibfnamefont {W.}~\bibnamefont {Schlotter}}, \bibinfo {author}
  {\bibfnamefont {J.}~\bibnamefont {Turner}}, \bibinfo {author} {\bibfnamefont
  {S.}~\bibnamefont {Johnson}}, \bibinfo {author} {\bibfnamefont
  {M.}~\bibnamefont {F{\"{o}}rst}}, \bibinfo {author} {\bibfnamefont
  {T.}~\bibnamefont {Sasagawa}}, \bibinfo {author} {\bibfnamefont
  {Y.}~\bibnamefont {Kung}}, \bibinfo {author} {\bibfnamefont {A.}~\bibnamefont
  {Sorini}}, \bibinfo {author} {\bibfnamefont {A.}~\bibnamefont {Kemper}},
  \bibinfo {author} {\bibfnamefont {B.}~\bibnamefont {Moritz}}, \bibinfo
  {author} {\bibfnamefont {T.}~\bibnamefont {Devereaux}}, \bibinfo {author}
  {\bibfnamefont {D.-H.}\ \bibnamefont {Lee}}, \bibinfo {author} {\bibfnamefont
  {Z.}~\bibnamefont {Shen}}, \ and\ \bibinfo {author} {\bibfnamefont
  {Z.}~\bibnamefont {Hussain}},\ }\href {\doibase 10.1038/ncomms1837}
  {\bibfield  {journal} {\bibinfo  {journal} {Nat. Commun.}\ }\textbf {\bibinfo
  {volume} {3}},\ \bibinfo {pages} {838} (\bibinfo {year} {2012})}\BibitemShut
  {NoStop}%
\bibitem [{\citenamefont {Caviglia}\ \emph {et~al.}(2013)\citenamefont
  {Caviglia}, \citenamefont {F{\"{o}}rst}, \citenamefont {Scherwitzl},
  \citenamefont {Khanna}, \citenamefont {Bromberger}, \citenamefont
  {Mankowsky}, \citenamefont {Singla}, \citenamefont {Chuang}, \citenamefont
  {Lee}, \citenamefont {Krupin}, \citenamefont {Schlotter}, \citenamefont
  {Turner}, \citenamefont {Dakovski}, \citenamefont {Minitti}, \citenamefont
  {Robinson}, \citenamefont {Scagnoli}, \citenamefont {Wilkins}, \citenamefont
  {Cavill}, \citenamefont {Gibert}, \citenamefont {Gariglio}, \citenamefont
  {Zubko}, \citenamefont {Triscone}, \citenamefont {Hill}, \citenamefont
  {Dhesi},\ and\ \citenamefont {Cavalleri}}]{Caviglia2013}%
  \BibitemOpen
  \bibfield  {author} {\bibinfo {author} {\bibfnamefont {A.~D.}\ \bibnamefont
  {Caviglia}}, \bibinfo {author} {\bibfnamefont {M.}~\bibnamefont
  {F{\"{o}}rst}}, \bibinfo {author} {\bibfnamefont {R.}~\bibnamefont
  {Scherwitzl}}, \bibinfo {author} {\bibfnamefont {V.}~\bibnamefont {Khanna}},
  \bibinfo {author} {\bibfnamefont {H.}~\bibnamefont {Bromberger}}, \bibinfo
  {author} {\bibfnamefont {R.}~\bibnamefont {Mankowsky}}, \bibinfo {author}
  {\bibfnamefont {R.}~\bibnamefont {Singla}}, \bibinfo {author} {\bibfnamefont
  {Y.-D.}\ \bibnamefont {Chuang}}, \bibinfo {author} {\bibfnamefont {W.~S.}\
  \bibnamefont {Lee}}, \bibinfo {author} {\bibfnamefont {O.}~\bibnamefont
  {Krupin}}, \bibinfo {author} {\bibfnamefont {W.~F.}\ \bibnamefont
  {Schlotter}}, \bibinfo {author} {\bibfnamefont {J.~J.}\ \bibnamefont
  {Turner}}, \bibinfo {author} {\bibfnamefont {G.~L.}\ \bibnamefont
  {Dakovski}}, \bibinfo {author} {\bibfnamefont {M.~P.}\ \bibnamefont
  {Minitti}}, \bibinfo {author} {\bibfnamefont {J.}~\bibnamefont {Robinson}},
  \bibinfo {author} {\bibfnamefont {V.}~\bibnamefont {Scagnoli}}, \bibinfo
  {author} {\bibfnamefont {S.~B.}\ \bibnamefont {Wilkins}}, \bibinfo {author}
  {\bibfnamefont {S.~A.}\ \bibnamefont {Cavill}}, \bibinfo {author}
  {\bibfnamefont {M.}~\bibnamefont {Gibert}}, \bibinfo {author} {\bibfnamefont
  {S.}~\bibnamefont {Gariglio}}, \bibinfo {author} {\bibfnamefont
  {P.}~\bibnamefont {Zubko}}, \bibinfo {author} {\bibfnamefont {J.-M.}\
  \bibnamefont {Triscone}}, \bibinfo {author} {\bibfnamefont {J.~P.}\
  \bibnamefont {Hill}}, \bibinfo {author} {\bibfnamefont {S.~S.}\ \bibnamefont
  {Dhesi}}, \ and\ \bibinfo {author} {\bibfnamefont {A.}~\bibnamefont
  {Cavalleri}},\ }\href {\doibase 10.1103/PhysRevB.88.220401} {\bibfield
  {journal} {\bibinfo  {journal} {Phys. Rev. B}\ }\textbf {\bibinfo {volume}
  {88}},\ \bibinfo {pages} {220401} (\bibinfo {year} {2013})}\BibitemShut
  {NoStop}%
\bibitem [{\citenamefont {Beaud}\ \emph {et~al.}(2014)\citenamefont {Beaud},
  \citenamefont {Caviezel}, \citenamefont {Mariager}, \citenamefont {Rettig},
  \citenamefont {Ingold}, \citenamefont {Dornes}, \citenamefont {Huang},
  \citenamefont {Johnson}, \citenamefont {Radovic}, \citenamefont {Huber},
  \citenamefont {Kubacka}, \citenamefont {Ferrer}, \citenamefont {Lemke},
  \citenamefont {Chollet}, \citenamefont {Zhu}, \citenamefont {Glownia},
  \citenamefont {Sikorski}, \citenamefont {Robert}, \citenamefont {Wadati},
  \citenamefont {Nakamura}, \citenamefont {Kawasaki}, \citenamefont {Tokura},
  \citenamefont {Johnson},\ and\ \citenamefont {Staub}}]{Beaud2014}%
  \BibitemOpen
  \bibfield  {author} {\bibinfo {author} {\bibfnamefont {P.}~\bibnamefont
  {Beaud}}, \bibinfo {author} {\bibfnamefont {A.}~\bibnamefont {Caviezel}},
  \bibinfo {author} {\bibfnamefont {S.~O.}\ \bibnamefont {Mariager}}, \bibinfo
  {author} {\bibfnamefont {L.}~\bibnamefont {Rettig}}, \bibinfo {author}
  {\bibfnamefont {G.}~\bibnamefont {Ingold}}, \bibinfo {author} {\bibfnamefont
  {C.}~\bibnamefont {Dornes}}, \bibinfo {author} {\bibfnamefont {S.-W.}\
  \bibnamefont {Huang}}, \bibinfo {author} {\bibfnamefont {J.~A.}\ \bibnamefont
  {Johnson}}, \bibinfo {author} {\bibfnamefont {M.}~\bibnamefont {Radovic}},
  \bibinfo {author} {\bibfnamefont {T.}~\bibnamefont {Huber}}, \bibinfo
  {author} {\bibfnamefont {T.}~\bibnamefont {Kubacka}}, \bibinfo {author}
  {\bibfnamefont {A.}~\bibnamefont {Ferrer}}, \bibinfo {author} {\bibfnamefont
  {H.~T.}\ \bibnamefont {Lemke}}, \bibinfo {author} {\bibfnamefont
  {M.}~\bibnamefont {Chollet}}, \bibinfo {author} {\bibfnamefont
  {D.}~\bibnamefont {Zhu}}, \bibinfo {author} {\bibfnamefont {J.~M.}\
  \bibnamefont {Glownia}}, \bibinfo {author} {\bibfnamefont {M.}~\bibnamefont
  {Sikorski}}, \bibinfo {author} {\bibfnamefont {A.}~\bibnamefont {Robert}},
  \bibinfo {author} {\bibfnamefont {H.}~\bibnamefont {Wadati}}, \bibinfo
  {author} {\bibfnamefont {M.}~\bibnamefont {Nakamura}}, \bibinfo {author}
  {\bibfnamefont {M.}~\bibnamefont {Kawasaki}}, \bibinfo {author}
  {\bibfnamefont {Y.}~\bibnamefont {Tokura}}, \bibinfo {author} {\bibfnamefont
  {S.~L.}\ \bibnamefont {Johnson}}, \ and\ \bibinfo {author} {\bibfnamefont
  {U.}~\bibnamefont {Staub}},\ }\href {\doibase 10.1038/nmat4046} {\bibfield
  {journal} {\bibinfo  {journal} {Nat. Mater.}\ }\textbf {\bibinfo {volume}
  {13}},\ \bibinfo {pages} {923} (\bibinfo {year} {2014})}\BibitemShut
  {NoStop}%
\bibitem [{\citenamefont {F{\"{o}}rst}\ \emph {et~al.}(2014)\citenamefont
  {F{\"{o}}rst}, \citenamefont {Tobey}, \citenamefont {Bromberger},
  \citenamefont {Wilkins}, \citenamefont {Khanna}, \citenamefont {Caviglia},
  \citenamefont {Chuang}, \citenamefont {Lee}, \citenamefont {Schlotter},
  \citenamefont {Turner}, \citenamefont {Minitti}, \citenamefont {Krupin},
  \citenamefont {Xu}, \citenamefont {Wen}, \citenamefont {Gu}, \citenamefont
  {Dhesi}, \citenamefont {Cavalleri},\ and\ \citenamefont {Hill}}]{Forst2014}%
  \BibitemOpen
  \bibfield  {author} {\bibinfo {author} {\bibfnamefont {M.}~\bibnamefont
  {F{\"{o}}rst}}, \bibinfo {author} {\bibfnamefont {R.~I.}\ \bibnamefont
  {Tobey}}, \bibinfo {author} {\bibfnamefont {H.}~\bibnamefont {Bromberger}},
  \bibinfo {author} {\bibfnamefont {S.~B.}\ \bibnamefont {Wilkins}}, \bibinfo
  {author} {\bibfnamefont {V.}~\bibnamefont {Khanna}}, \bibinfo {author}
  {\bibfnamefont {A.~D.}\ \bibnamefont {Caviglia}}, \bibinfo {author}
  {\bibfnamefont {Y.-D.}\ \bibnamefont {Chuang}}, \bibinfo {author}
  {\bibfnamefont {W.~S.}\ \bibnamefont {Lee}}, \bibinfo {author} {\bibfnamefont
  {W.~F.}\ \bibnamefont {Schlotter}}, \bibinfo {author} {\bibfnamefont {J.~J.}\
  \bibnamefont {Turner}}, \bibinfo {author} {\bibfnamefont {M.~P.}\
  \bibnamefont {Minitti}}, \bibinfo {author} {\bibfnamefont {O.}~\bibnamefont
  {Krupin}}, \bibinfo {author} {\bibfnamefont {Z.~J.}\ \bibnamefont {Xu}},
  \bibinfo {author} {\bibfnamefont {J.~S.}\ \bibnamefont {Wen}}, \bibinfo
  {author} {\bibfnamefont {G.~D.}\ \bibnamefont {Gu}}, \bibinfo {author}
  {\bibfnamefont {S.~S.}\ \bibnamefont {Dhesi}}, \bibinfo {author}
  {\bibfnamefont {A.}~\bibnamefont {Cavalleri}}, \ and\ \bibinfo {author}
  {\bibfnamefont {J.~P.}\ \bibnamefont {Hill}},\ }\href {\doibase
  10.1103/PhysRevLett.112.157002} {\bibfield  {journal} {\bibinfo  {journal}
  {Phys. Rev. Lett.}\ }\textbf {\bibinfo {volume} {112}},\ \bibinfo {pages}
  {157002} (\bibinfo {year} {2014})}\BibitemShut {NoStop}%
\bibitem [{\citenamefont {Forst}\ \emph {et~al.}(2015)\citenamefont {Forst},
  \citenamefont {Caviglia}, \citenamefont {Scherwitzl}, \citenamefont
  {Mankowsky}, \citenamefont {Zubko}, \citenamefont {Khanna}, \citenamefont
  {Bromberger}, \citenamefont {Wilkins}, \citenamefont {Chuang}, \citenamefont
  {Lee}, \citenamefont {Schlotter}, \citenamefont {Turner}, \citenamefont
  {Dakovski}, \citenamefont {Minitti}, \citenamefont {Robinson}, \citenamefont
  {Clark}, \citenamefont {Jaksch}, \citenamefont {Triscone}, \citenamefont
  {Hill}, \citenamefont {Dhesi},\ and\ \citenamefont {Cavalleri}}]{Forst2015}%
  \BibitemOpen
  \bibfield  {author} {\bibinfo {author} {\bibfnamefont {M.}~\bibnamefont
  {Forst}}, \bibinfo {author} {\bibfnamefont {A.~D.}\ \bibnamefont {Caviglia}},
  \bibinfo {author} {\bibfnamefont {R.}~\bibnamefont {Scherwitzl}}, \bibinfo
  {author} {\bibfnamefont {R.}~\bibnamefont {Mankowsky}}, \bibinfo {author}
  {\bibfnamefont {P.}~\bibnamefont {Zubko}}, \bibinfo {author} {\bibfnamefont
  {V.}~\bibnamefont {Khanna}}, \bibinfo {author} {\bibfnamefont
  {H.}~\bibnamefont {Bromberger}}, \bibinfo {author} {\bibfnamefont {S.~B.}\
  \bibnamefont {Wilkins}}, \bibinfo {author} {\bibfnamefont {Y.~D.}\
  \bibnamefont {Chuang}}, \bibinfo {author} {\bibfnamefont {W.~S.}\
  \bibnamefont {Lee}}, \bibinfo {author} {\bibfnamefont {W.~F.}\ \bibnamefont
  {Schlotter}}, \bibinfo {author} {\bibfnamefont {J.~J.}\ \bibnamefont
  {Turner}}, \bibinfo {author} {\bibfnamefont {G.~L.}\ \bibnamefont
  {Dakovski}}, \bibinfo {author} {\bibfnamefont {M.~P.}\ \bibnamefont
  {Minitti}}, \bibinfo {author} {\bibfnamefont {J.}~\bibnamefont {Robinson}},
  \bibinfo {author} {\bibfnamefont {S.~R.}\ \bibnamefont {Clark}}, \bibinfo
  {author} {\bibfnamefont {D.}~\bibnamefont {Jaksch}}, \bibinfo {author}
  {\bibfnamefont {J.~M.}\ \bibnamefont {Triscone}}, \bibinfo {author}
  {\bibfnamefont {J.~P.}\ \bibnamefont {Hill}}, \bibinfo {author}
  {\bibfnamefont {S.~S.}\ \bibnamefont {Dhesi}}, \ and\ \bibinfo {author}
  {\bibfnamefont {A.}~\bibnamefont {Cavalleri}},\ }\href {\doibase
  10.1038/nmat4341} {\bibfield  {journal} {\bibinfo  {journal} {Nat. Mater.}\
  }\textbf {\bibinfo {volume} {14}},\ \bibinfo {pages} {883} (\bibinfo {year}
  {2015})}\BibitemShut {NoStop}%
\bibitem [{\citenamefont {Tsuyama}\ \emph {et~al.}(2016)\citenamefont
  {Tsuyama}, \citenamefont {Chakraverty}, \citenamefont {Macke}, \citenamefont
  {Pontius}, \citenamefont {Sch{\"{u}}{\ss}ler-Langeheine}, \citenamefont
  {Hwang}, \citenamefont {Tokura},\ and\ \citenamefont {Wadati}}]{Tsuyama2016}%
  \BibitemOpen
  \bibfield  {author} {\bibinfo {author} {\bibfnamefont {T.}~\bibnamefont
  {Tsuyama}}, \bibinfo {author} {\bibfnamefont {S.}~\bibnamefont
  {Chakraverty}}, \bibinfo {author} {\bibfnamefont {S.}~\bibnamefont {Macke}},
  \bibinfo {author} {\bibfnamefont {N.}~\bibnamefont {Pontius}}, \bibinfo
  {author} {\bibfnamefont {C.}~\bibnamefont {Sch{\"{u}}{\ss}ler-Langeheine}},
  \bibinfo {author} {\bibfnamefont {H.~Y.}\ \bibnamefont {Hwang}}, \bibinfo
  {author} {\bibfnamefont {Y.}~\bibnamefont {Tokura}}, \ and\ \bibinfo {author}
  {\bibfnamefont {H.}~\bibnamefont {Wadati}},\ }\href {\doibase
  10.1103/PhysRevLett.116.256402} {\bibfield  {journal} {\bibinfo  {journal}
  {Phys. Rev. Lett.}\ }\textbf {\bibinfo {volume} {116}},\ \bibinfo {pages}
  {256402} (\bibinfo {year} {2016})}\BibitemShut {NoStop}%
\bibitem [{\citenamefont {Thielemann-K{\"{u}}hn}\ \emph
  {et~al.}(2017)\citenamefont {Thielemann-K{\"{u}}hn}, \citenamefont {Schick},
  \citenamefont {Pontius}, \citenamefont {Trabant}, \citenamefont {Mitzner},
  \citenamefont {Holldack}, \citenamefont {Zabel}, \citenamefont
  {F{\"{o}}hlisch},\ and\ \citenamefont
  {Sch{\"{u}}{\ss}ler-Langeheine}}]{Thielemann-Kuhn2017a}%
  \BibitemOpen
  \bibfield  {author} {\bibinfo {author} {\bibfnamefont {N.}~\bibnamefont
  {Thielemann-K{\"{u}}hn}}, \bibinfo {author} {\bibfnamefont {D.}~\bibnamefont
  {Schick}}, \bibinfo {author} {\bibfnamefont {N.}~\bibnamefont {Pontius}},
  \bibinfo {author} {\bibfnamefont {C.}~\bibnamefont {Trabant}}, \bibinfo
  {author} {\bibfnamefont {R.}~\bibnamefont {Mitzner}}, \bibinfo {author}
  {\bibfnamefont {K.}~\bibnamefont {Holldack}}, \bibinfo {author}
  {\bibfnamefont {H.}~\bibnamefont {Zabel}}, \bibinfo {author} {\bibfnamefont
  {A.}~\bibnamefont {F{\"{o}}hlisch}}, \ and\ \bibinfo {author} {\bibfnamefont
  {C.}~\bibnamefont {Sch{\"{u}}{\ss}ler-Langeheine}},\ }\href {\doibase
  10.1103/PhysRevLett.119.197202} {\bibfield  {journal} {\bibinfo  {journal}
  {Phys. Rev. Lett.}\ }\textbf {\bibinfo {volume} {119}},\ \bibinfo {pages}
  {197202} (\bibinfo {year} {2017})}\BibitemShut {NoStop}%
\bibitem [{\citenamefont {Imada}\ \emph {et~al.}(1998)\citenamefont {Imada},
  \citenamefont {Fujimori},\ and\ \citenamefont {Tokura}}]{Imada1998}%
  \BibitemOpen
  \bibfield  {author} {\bibinfo {author} {\bibfnamefont {M.}~\bibnamefont
  {Imada}}, \bibinfo {author} {\bibfnamefont {A.}~\bibnamefont {Fujimori}}, \
  and\ \bibinfo {author} {\bibfnamefont {Y.}~\bibnamefont {Tokura}},\ }\href
  {\doibase 10.1103/RevModPhys.70.1039} {\bibfield  {journal} {\bibinfo
  {journal} {Rev. Mod. Phys.}\ }\textbf {\bibinfo {volume} {70}},\ \bibinfo
  {pages} {1039} (\bibinfo {year} {1998})}\BibitemShut {NoStop}%
\bibitem [{\citenamefont {Wadati}\ \emph {et~al.}(2005)\citenamefont {Wadati},
  \citenamefont {Kobayashi}, \citenamefont {Kumigashira}, \citenamefont
  {Okazaki}, \citenamefont {Mizokawa}, \citenamefont {Fujimori}, \citenamefont
  {Horiba}, \citenamefont {Oshima}, \citenamefont {Hamada}, \citenamefont
  {Lippmaa}, \citenamefont {Kawasaki},\ and\ \citenamefont
  {Koinuma}}]{Wadati2005}%
  \BibitemOpen
  \bibfield  {author} {\bibinfo {author} {\bibfnamefont {H.}~\bibnamefont
  {Wadati}}, \bibinfo {author} {\bibfnamefont {D.}~\bibnamefont {Kobayashi}},
  \bibinfo {author} {\bibfnamefont {H.}~\bibnamefont {Kumigashira}}, \bibinfo
  {author} {\bibfnamefont {K.}~\bibnamefont {Okazaki}}, \bibinfo {author}
  {\bibfnamefont {T.}~\bibnamefont {Mizokawa}}, \bibinfo {author}
  {\bibfnamefont {A.}~\bibnamefont {Fujimori}}, \bibinfo {author}
  {\bibfnamefont {K.}~\bibnamefont {Horiba}}, \bibinfo {author} {\bibfnamefont
  {M.}~\bibnamefont {Oshima}}, \bibinfo {author} {\bibfnamefont
  {N.}~\bibnamefont {Hamada}}, \bibinfo {author} {\bibfnamefont
  {M.}~\bibnamefont {Lippmaa}}, \bibinfo {author} {\bibfnamefont
  {M.}~\bibnamefont {Kawasaki}}, \ and\ \bibinfo {author} {\bibfnamefont
  {H.}~\bibnamefont {Koinuma}},\ }\href {\doibase 10.1103/PhysRevB.71.035108}
  {\bibfield  {journal} {\bibinfo  {journal} {Phys. Rev. B}\ }\textbf {\bibinfo
  {volume} {71}},\ \bibinfo {pages} {035108} (\bibinfo {year}
  {2005})}\BibitemShut {NoStop}%
\bibitem [{\citenamefont {Sichel-Tissot}\ \emph {et~al.}(2013)\citenamefont
  {Sichel-Tissot}, \citenamefont {Devlin}, \citenamefont {Ryan}, \citenamefont
  {Kim},\ and\ \citenamefont {May}}]{Sichel-Tissot2013}%
  \BibitemOpen
  \bibfield  {author} {\bibinfo {author} {\bibfnamefont {R.~J.}\ \bibnamefont
  {Sichel-Tissot}}, \bibinfo {author} {\bibfnamefont {R.~C.}\ \bibnamefont
  {Devlin}}, \bibinfo {author} {\bibfnamefont {P.~J.}\ \bibnamefont {Ryan}},
  \bibinfo {author} {\bibfnamefont {J.-w.}\ \bibnamefont {Kim}}, \ and\
  \bibinfo {author} {\bibfnamefont {S.~J.}\ \bibnamefont {May}},\ }\href
  {\doibase 10.1063/1.4833276} {\bibfield  {journal} {\bibinfo  {journal}
  {Appl. Phys. Lett.}\ }\textbf {\bibinfo {volume} {103}},\ \bibinfo {pages}
  {212905} (\bibinfo {year} {2013})}\BibitemShut {NoStop}%
\bibitem [{\citenamefont {Abbate}\ \emph {et~al.}(1992)\citenamefont {Abbate},
  \citenamefont {de~Groot}, \citenamefont {Fuggle}, \citenamefont {Fujimori},
  \citenamefont {Strebel}, \citenamefont {Lopez}, \citenamefont {Domke},
  \citenamefont {Kaindl}, \citenamefont {Sawatzky}, \citenamefont {Takano},
  \citenamefont {Takeda}, \citenamefont {Eisaki},\ and\ \citenamefont
  {Uchida}}]{Abbate1992}%
  \BibitemOpen
  \bibfield  {author} {\bibinfo {author} {\bibfnamefont {M.}~\bibnamefont
  {Abbate}}, \bibinfo {author} {\bibfnamefont {F.~M.~F.}\ \bibnamefont
  {de~Groot}}, \bibinfo {author} {\bibfnamefont {J.~C.}\ \bibnamefont
  {Fuggle}}, \bibinfo {author} {\bibfnamefont {A.}~\bibnamefont {Fujimori}},
  \bibinfo {author} {\bibfnamefont {O.}~\bibnamefont {Strebel}}, \bibinfo
  {author} {\bibfnamefont {F.}~\bibnamefont {Lopez}}, \bibinfo {author}
  {\bibfnamefont {M.}~\bibnamefont {Domke}}, \bibinfo {author} {\bibfnamefont
  {G.}~\bibnamefont {Kaindl}}, \bibinfo {author} {\bibfnamefont {G.~A.}\
  \bibnamefont {Sawatzky}}, \bibinfo {author} {\bibfnamefont {M.}~\bibnamefont
  {Takano}}, \bibinfo {author} {\bibfnamefont {Y.}~\bibnamefont {Takeda}},
  \bibinfo {author} {\bibfnamefont {H.}~\bibnamefont {Eisaki}}, \ and\ \bibinfo
  {author} {\bibfnamefont {S.}~\bibnamefont {Uchida}},\ }\href {\doibase
  10.1103/PhysRevB.46.4511} {\bibfield  {journal} {\bibinfo  {journal} {Phys.
  Rev. B}\ }\textbf {\bibinfo {volume} {46}},\ \bibinfo {pages} {4511}
  (\bibinfo {year} {1992})}\BibitemShut {NoStop}%
\bibitem [{\citenamefont {Takeda}\ \emph {et~al.}(1972)\citenamefont {Takeda},
  \citenamefont {Yamaguchi},\ and\ \citenamefont {Watanabe}}]{Takeda1972}%
  \BibitemOpen
  \bibfield  {author} {\bibinfo {author} {\bibfnamefont {T.}~\bibnamefont
  {Takeda}}, \bibinfo {author} {\bibfnamefont {Y.}~\bibnamefont {Yamaguchi}}, \
  and\ \bibinfo {author} {\bibfnamefont {H.}~\bibnamefont {Watanabe}},\ }\href
  {\doibase 10.1143/JPSJ.33.967} {\bibfield  {journal} {\bibinfo  {journal} {J.
  Phys. Soc. Jpn.}\ }\textbf {\bibinfo {volume} {33}},\ \bibinfo {pages} {967}
  (\bibinfo {year} {1972})}\BibitemShut {NoStop}%
\bibitem [{\citenamefont {Chakraverty}\ \emph {et~al.}(2013)\citenamefont
  {Chakraverty}, \citenamefont {Matsuda}, \citenamefont {Wadati}, \citenamefont
  {Okamoto}, \citenamefont {Yamasaki}, \citenamefont {Nakao}, \citenamefont
  {Murakami}, \citenamefont {Ishiwata}, \citenamefont {Kawasaki}, \citenamefont
  {Taguchi}, \citenamefont {Tokura},\ and\ \citenamefont
  {Hwang}}]{Chakraverty2013}%
  \BibitemOpen
  \bibfield  {author} {\bibinfo {author} {\bibfnamefont {S.}~\bibnamefont
  {Chakraverty}}, \bibinfo {author} {\bibfnamefont {T.}~\bibnamefont
  {Matsuda}}, \bibinfo {author} {\bibfnamefont {H.}~\bibnamefont {Wadati}},
  \bibinfo {author} {\bibfnamefont {J.}~\bibnamefont {Okamoto}}, \bibinfo
  {author} {\bibfnamefont {Y.}~\bibnamefont {Yamasaki}}, \bibinfo {author}
  {\bibfnamefont {H.}~\bibnamefont {Nakao}}, \bibinfo {author} {\bibfnamefont
  {Y.}~\bibnamefont {Murakami}}, \bibinfo {author} {\bibfnamefont
  {S.}~\bibnamefont {Ishiwata}}, \bibinfo {author} {\bibfnamefont
  {M.}~\bibnamefont {Kawasaki}}, \bibinfo {author} {\bibfnamefont
  {Y.}~\bibnamefont {Taguchi}}, \bibinfo {author} {\bibfnamefont
  {Y.}~\bibnamefont {Tokura}}, \ and\ \bibinfo {author} {\bibfnamefont {H.~Y.}\
  \bibnamefont {Hwang}},\ }\href {\doibase 10.1103/PhysRevB.88.220405}
  {\bibfield  {journal} {\bibinfo  {journal} {Phys. Rev. B}\ }\textbf {\bibinfo
  {volume} {88}},\ \bibinfo {pages} {220405} (\bibinfo {year}
  {2013})}\BibitemShut {NoStop}%
\bibitem [{\citenamefont {Rogge}\ \emph {et~al.}(2019)\citenamefont {Rogge},
  \citenamefont {Green}, \citenamefont {Sutarto},\ and\ \citenamefont
  {May}}]{Rogge2019}%
  \BibitemOpen
  \bibfield  {author} {\bibinfo {author} {\bibfnamefont {P.~C.}\ \bibnamefont
  {Rogge}}, \bibinfo {author} {\bibfnamefont {R.~J.}\ \bibnamefont {Green}},
  \bibinfo {author} {\bibfnamefont {R.}~\bibnamefont {Sutarto}}, \ and\
  \bibinfo {author} {\bibfnamefont {S.~J.}\ \bibnamefont {May}},\ }\href
  {\doibase 10.1103/PhysRevMaterials.3.084404} {\bibfield  {journal} {\bibinfo
  {journal} {Phys. Rev. Mater.}\ }\textbf {\bibinfo {volume} {3}},\ \bibinfo
  {pages} {084404} (\bibinfo {year} {2019})}\BibitemShut {NoStop}%
\bibitem [{\citenamefont {Ishiwata}\ \emph {et~al.}(2011)\citenamefont
  {Ishiwata}, \citenamefont {Tokunaga}, \citenamefont {Kaneko}, \citenamefont
  {Okuyama}, \citenamefont {Tokunaga}, \citenamefont {Wakimoto}, \citenamefont
  {Kakurai}, \citenamefont {Arima}, \citenamefont {Taguchi},\ and\
  \citenamefont {Tokura}}]{Ishiwata2011}%
  \BibitemOpen
  \bibfield  {author} {\bibinfo {author} {\bibfnamefont {S.}~\bibnamefont
  {Ishiwata}}, \bibinfo {author} {\bibfnamefont {M.}~\bibnamefont {Tokunaga}},
  \bibinfo {author} {\bibfnamefont {Y.}~\bibnamefont {Kaneko}}, \bibinfo
  {author} {\bibfnamefont {D.}~\bibnamefont {Okuyama}}, \bibinfo {author}
  {\bibfnamefont {Y.}~\bibnamefont {Tokunaga}}, \bibinfo {author}
  {\bibfnamefont {S.}~\bibnamefont {Wakimoto}}, \bibinfo {author}
  {\bibfnamefont {K.}~\bibnamefont {Kakurai}}, \bibinfo {author} {\bibfnamefont
  {T.}~\bibnamefont {Arima}}, \bibinfo {author} {\bibfnamefont
  {Y.}~\bibnamefont {Taguchi}}, \ and\ \bibinfo {author} {\bibfnamefont
  {Y.}~\bibnamefont {Tokura}},\ }\href {\doibase 10.1103/PhysRevB.84.054427}
  {\bibfield  {journal} {\bibinfo  {journal} {Phys. Rev. B}\ }\textbf {\bibinfo
  {volume} {84}},\ \bibinfo {pages} {054427} (\bibinfo {year}
  {2011})}\BibitemShut {NoStop}%
\bibitem [{\citenamefont {Ishiwata}\ \emph {et~al.}(2020)\citenamefont
  {Ishiwata}, \citenamefont {Nakajima}, \citenamefont {Kim}, \citenamefont
  {Inosov}, \citenamefont {Kanazawa}, \citenamefont {White}, \citenamefont
  {Gavilano}, \citenamefont {Georgii}, \citenamefont {Seemann}, \citenamefont
  {Brandl}, \citenamefont {Manuel}, \citenamefont {Khalyavin}, \citenamefont
  {Seki}, \citenamefont {Tokunaga}, \citenamefont {Kinoshita}, \citenamefont
  {Long}, \citenamefont {Kaneko}, \citenamefont {Taguchi}, \citenamefont
  {Arima}, \citenamefont {Keimer},\ and\ \citenamefont
  {Tokura}}]{Ishiwata2020}%
  \BibitemOpen
  \bibfield  {author} {\bibinfo {author} {\bibfnamefont {S.}~\bibnamefont
  {Ishiwata}}, \bibinfo {author} {\bibfnamefont {T.}~\bibnamefont {Nakajima}},
  \bibinfo {author} {\bibfnamefont {J.-H.}\ \bibnamefont {Kim}}, \bibinfo
  {author} {\bibfnamefont {D.~S.}\ \bibnamefont {Inosov}}, \bibinfo {author}
  {\bibfnamefont {N.}~\bibnamefont {Kanazawa}}, \bibinfo {author}
  {\bibfnamefont {J.~S.}\ \bibnamefont {White}}, \bibinfo {author}
  {\bibfnamefont {J.~L.}\ \bibnamefont {Gavilano}}, \bibinfo {author}
  {\bibfnamefont {R.}~\bibnamefont {Georgii}}, \bibinfo {author} {\bibfnamefont
  {K.~M.}\ \bibnamefont {Seemann}}, \bibinfo {author} {\bibfnamefont
  {G.}~\bibnamefont {Brandl}}, \bibinfo {author} {\bibfnamefont
  {P.}~\bibnamefont {Manuel}}, \bibinfo {author} {\bibfnamefont {D.~D.}\
  \bibnamefont {Khalyavin}}, \bibinfo {author} {\bibfnamefont {S.}~\bibnamefont
  {Seki}}, \bibinfo {author} {\bibfnamefont {Y.}~\bibnamefont {Tokunaga}},
  \bibinfo {author} {\bibfnamefont {M.}~\bibnamefont {Kinoshita}}, \bibinfo
  {author} {\bibfnamefont {Y.~W.}\ \bibnamefont {Long}}, \bibinfo {author}
  {\bibfnamefont {Y.}~\bibnamefont {Kaneko}}, \bibinfo {author} {\bibfnamefont
  {Y.}~\bibnamefont {Taguchi}}, \bibinfo {author} {\bibfnamefont
  {T.}~\bibnamefont {Arima}}, \bibinfo {author} {\bibfnamefont
  {B.}~\bibnamefont {Keimer}}, \ and\ \bibinfo {author} {\bibfnamefont
  {Y.}~\bibnamefont {Tokura}},\ }\href {\doibase 10.1103/PhysRevB.101.134406}
  {\bibfield  {journal} {\bibinfo  {journal} {Phys. Rev. B}\ }\textbf {\bibinfo
  {volume} {101}},\ \bibinfo {pages} {134406} (\bibinfo {year}
  {2020})}\BibitemShut {NoStop}%
\bibitem [{\citenamefont {Fink}\ \emph {et~al.}(2013)\citenamefont {Fink},
  \citenamefont {Schierle}, \citenamefont {Weschke},\ and\ \citenamefont
  {Geck}}]{Fink2013}%
  \BibitemOpen
  \bibfield  {author} {\bibinfo {author} {\bibfnamefont {J.}~\bibnamefont
  {Fink}}, \bibinfo {author} {\bibfnamefont {E.}~\bibnamefont {Schierle}},
  \bibinfo {author} {\bibfnamefont {E.}~\bibnamefont {Weschke}}, \ and\
  \bibinfo {author} {\bibfnamefont {J.}~\bibnamefont {Geck}},\ }\href {\doibase
  10.1088/0034-4885/76/5/056502} {\bibfield  {journal} {\bibinfo  {journal}
  {Reports Prog. Phys.}\ }\textbf {\bibinfo {volume} {76}},\ \bibinfo {pages}
  {056502} (\bibinfo {year} {2013})}\BibitemShut {NoStop}%
\bibitem [{\citenamefont {Sch{\"{u}}{\ss}ler-Langeheine}\ \emph
  {et~al.}(2001)\citenamefont {Sch{\"{u}}{\ss}ler-Langeheine}, \citenamefont
  {Weschke}, \citenamefont {Grigoriev}, \citenamefont {Ott}, \citenamefont
  {Meier}, \citenamefont {Vyalikh}, \citenamefont {Mazumdar}, \citenamefont
  {Sutter}, \citenamefont {Abernathy}, \citenamefont {Gr{\"{u}}bel},\ and\
  \citenamefont {Kaindl}}]{Schuessler-Langeheine2001}%
  \BibitemOpen
  \bibfield  {author} {\bibinfo {author} {\bibfnamefont {C.}~\bibnamefont
  {Sch{\"{u}}{\ss}ler-Langeheine}}, \bibinfo {author} {\bibfnamefont
  {E.}~\bibnamefont {Weschke}}, \bibinfo {author} {\bibfnamefont
  {A.}~\bibnamefont {Grigoriev}}, \bibinfo {author} {\bibfnamefont
  {H.}~\bibnamefont {Ott}}, \bibinfo {author} {\bibfnamefont {R.}~\bibnamefont
  {Meier}}, \bibinfo {author} {\bibfnamefont {D.}~\bibnamefont {Vyalikh}},
  \bibinfo {author} {\bibfnamefont {C.}~\bibnamefont {Mazumdar}}, \bibinfo
  {author} {\bibfnamefont {C.}~\bibnamefont {Sutter}}, \bibinfo {author}
  {\bibfnamefont {D.}~\bibnamefont {Abernathy}}, \bibinfo {author}
  {\bibfnamefont {G.}~\bibnamefont {Gr{\"{u}}bel}}, \ and\ \bibinfo {author}
  {\bibfnamefont {G.}~\bibnamefont {Kaindl}},\ }\href {\doibase
  10.1016/S0368-2048(00)00318-2} {\bibfield  {journal} {\bibinfo  {journal} {J.
  Electron Spectros. Relat. Phenomena}\ }\textbf {\bibinfo {volume}
  {114-116}},\ \bibinfo {pages} {953} (\bibinfo {year} {2001})}\BibitemShut
  {NoStop}%
\bibitem [{\citenamefont {Zhou}\ \emph {et~al.}(2011)\citenamefont {Zhou},
  \citenamefont {Zhu}, \citenamefont {Langner}, \citenamefont {Chuang},
  \citenamefont {Yu}, \citenamefont {Yang}, \citenamefont {{Cruz Gonzalez}},
  \citenamefont {Tahir}, \citenamefont {Rini}, \citenamefont {Chu},
  \citenamefont {Ramesh}, \citenamefont {Lee}, \citenamefont {Tomioka},
  \citenamefont {Tokura}, \citenamefont {Hussain},\ and\ \citenamefont
  {Schoenlein}}]{Zhou2011}%
  \BibitemOpen
  \bibfield  {author} {\bibinfo {author} {\bibfnamefont {S.~Y.}\ \bibnamefont
  {Zhou}}, \bibinfo {author} {\bibfnamefont {Y.}~\bibnamefont {Zhu}}, \bibinfo
  {author} {\bibfnamefont {M.~C.}\ \bibnamefont {Langner}}, \bibinfo {author}
  {\bibfnamefont {Y.-D.}\ \bibnamefont {Chuang}}, \bibinfo {author}
  {\bibfnamefont {P.}~\bibnamefont {Yu}}, \bibinfo {author} {\bibfnamefont
  {W.~L.}\ \bibnamefont {Yang}}, \bibinfo {author} {\bibfnamefont {A.~G.}\
  \bibnamefont {{Cruz Gonzalez}}}, \bibinfo {author} {\bibfnamefont
  {N.}~\bibnamefont {Tahir}}, \bibinfo {author} {\bibfnamefont
  {M.}~\bibnamefont {Rini}}, \bibinfo {author} {\bibfnamefont {Y.-H.}\
  \bibnamefont {Chu}}, \bibinfo {author} {\bibfnamefont {R.}~\bibnamefont
  {Ramesh}}, \bibinfo {author} {\bibfnamefont {D.-H.}\ \bibnamefont {Lee}},
  \bibinfo {author} {\bibfnamefont {Y.}~\bibnamefont {Tomioka}}, \bibinfo
  {author} {\bibfnamefont {Y.}~\bibnamefont {Tokura}}, \bibinfo {author}
  {\bibfnamefont {Z.}~\bibnamefont {Hussain}}, \ and\ \bibinfo {author}
  {\bibfnamefont {R.~W.}\ \bibnamefont {Schoenlein}},\ }\href {\doibase
  10.1103/PhysRevLett.106.186404} {\bibfield  {journal} {\bibinfo  {journal}
  {Phys. Rev. Lett.}\ }\textbf {\bibinfo {volume} {106}},\ \bibinfo {pages}
  {186404} (\bibinfo {year} {2011})}\BibitemShut {NoStop}%
\bibitem [{\citenamefont {Partzsch}\ \emph {et~al.}(2011)\citenamefont
  {Partzsch}, \citenamefont {Wilkins}, \citenamefont {Hill}, \citenamefont
  {Schierle}, \citenamefont {Weschke}, \citenamefont {Souptel}, \citenamefont
  {B{\"{u}}chner},\ and\ \citenamefont {Geck}}]{Partzsch2011}%
  \BibitemOpen
  \bibfield  {author} {\bibinfo {author} {\bibfnamefont {S.}~\bibnamefont
  {Partzsch}}, \bibinfo {author} {\bibfnamefont {S.~B.}\ \bibnamefont
  {Wilkins}}, \bibinfo {author} {\bibfnamefont {J.~P.}\ \bibnamefont {Hill}},
  \bibinfo {author} {\bibfnamefont {E.}~\bibnamefont {Schierle}}, \bibinfo
  {author} {\bibfnamefont {E.}~\bibnamefont {Weschke}}, \bibinfo {author}
  {\bibfnamefont {D.}~\bibnamefont {Souptel}}, \bibinfo {author} {\bibfnamefont
  {B.}~\bibnamefont {B{\"{u}}chner}}, \ and\ \bibinfo {author} {\bibfnamefont
  {J.}~\bibnamefont {Geck}},\ }\href {\doibase 10.1103/PhysRevLett.107.057201}
  {\bibfield  {journal} {\bibinfo  {journal} {Phys. Rev. Lett.}\ }\textbf
  {\bibinfo {volume} {107}},\ \bibinfo {pages} {057201} (\bibinfo {year}
  {2011})}\BibitemShut {NoStop}%
\bibitem [{\citenamefont {Wadati}\ \emph {et~al.}(2012)\citenamefont {Wadati},
  \citenamefont {Okamoto}, \citenamefont {Garganourakis}, \citenamefont
  {Scagnoli}, \citenamefont {Staub}, \citenamefont {Yamasaki}, \citenamefont
  {Nakao}, \citenamefont {Murakami}, \citenamefont {Mochizuki}, \citenamefont
  {Nakamura}, \citenamefont {Kawasaki},\ and\ \citenamefont
  {Tokura}}]{Wadati2012}%
  \BibitemOpen
  \bibfield  {author} {\bibinfo {author} {\bibfnamefont {H.}~\bibnamefont
  {Wadati}}, \bibinfo {author} {\bibfnamefont {J.}~\bibnamefont {Okamoto}},
  \bibinfo {author} {\bibfnamefont {M.}~\bibnamefont {Garganourakis}}, \bibinfo
  {author} {\bibfnamefont {V.}~\bibnamefont {Scagnoli}}, \bibinfo {author}
  {\bibfnamefont {U.}~\bibnamefont {Staub}}, \bibinfo {author} {\bibfnamefont
  {Y.}~\bibnamefont {Yamasaki}}, \bibinfo {author} {\bibfnamefont
  {H.}~\bibnamefont {Nakao}}, \bibinfo {author} {\bibfnamefont
  {Y.}~\bibnamefont {Murakami}}, \bibinfo {author} {\bibfnamefont
  {M.}~\bibnamefont {Mochizuki}}, \bibinfo {author} {\bibfnamefont
  {M.}~\bibnamefont {Nakamura}}, \bibinfo {author} {\bibfnamefont
  {M.}~\bibnamefont {Kawasaki}}, \ and\ \bibinfo {author} {\bibfnamefont
  {Y.}~\bibnamefont {Tokura}},\ }\href {\doibase
  10.1103/PhysRevLett.108.047203} {\bibfield  {journal} {\bibinfo  {journal}
  {Phys. Rev. Lett.}\ }\textbf {\bibinfo {volume} {108}},\ \bibinfo {pages}
  {047203} (\bibinfo {year} {2012})}\BibitemShut {NoStop}%
\bibitem [{\citenamefont {Matsuda}\ \emph {et~al.}(2015)\citenamefont
  {Matsuda}, \citenamefont {Partzsch}, \citenamefont {Tsuyama}, \citenamefont
  {Schierle}, \citenamefont {Weschke}, \citenamefont {Geck}, \citenamefont
  {Saito}, \citenamefont {Ishiwata}, \citenamefont {Tokura},\ and\
  \citenamefont {Wadati}}]{Matsuda2015a}%
  \BibitemOpen
  \bibfield  {author} {\bibinfo {author} {\bibfnamefont {T.}~\bibnamefont
  {Matsuda}}, \bibinfo {author} {\bibfnamefont {S.}~\bibnamefont {Partzsch}},
  \bibinfo {author} {\bibfnamefont {T.}~\bibnamefont {Tsuyama}}, \bibinfo
  {author} {\bibfnamefont {E.}~\bibnamefont {Schierle}}, \bibinfo {author}
  {\bibfnamefont {E.}~\bibnamefont {Weschke}}, \bibinfo {author} {\bibfnamefont
  {J.}~\bibnamefont {Geck}}, \bibinfo {author} {\bibfnamefont {T.}~\bibnamefont
  {Saito}}, \bibinfo {author} {\bibfnamefont {S.}~\bibnamefont {Ishiwata}},
  \bibinfo {author} {\bibfnamefont {Y.}~\bibnamefont {Tokura}}, \ and\ \bibinfo
  {author} {\bibfnamefont {H.}~\bibnamefont {Wadati}},\ }\href {\doibase
  10.1103/PhysRevLett.114.236403} {\bibfield  {journal} {\bibinfo  {journal}
  {Phys. Rev. Lett.}\ }\textbf {\bibinfo {volume} {114}},\ \bibinfo {pages}
  {236403} (\bibinfo {year} {2015})}\BibitemShut {NoStop}%
\bibitem [{\citenamefont {Yamamoto}\ \emph {et~al.}(2018)\citenamefont
  {Yamamoto}, \citenamefont {Hirata}, \citenamefont {Horio}, \citenamefont
  {Yokoyama}, \citenamefont {Takubo}, \citenamefont {Minohara}, \citenamefont
  {Kumigashira}, \citenamefont {Yamasaki}, \citenamefont {Nakao}, \citenamefont
  {Murakami}, \citenamefont {Fujimori},\ and\ \citenamefont
  {Wadati}}]{Yamamoto2018}%
  \BibitemOpen
  \bibfield  {author} {\bibinfo {author} {\bibfnamefont {K.}~\bibnamefont
  {Yamamoto}}, \bibinfo {author} {\bibfnamefont {Y.}~\bibnamefont {Hirata}},
  \bibinfo {author} {\bibfnamefont {M.}~\bibnamefont {Horio}}, \bibinfo
  {author} {\bibfnamefont {Y.}~\bibnamefont {Yokoyama}}, \bibinfo {author}
  {\bibfnamefont {K.}~\bibnamefont {Takubo}}, \bibinfo {author} {\bibfnamefont
  {M.}~\bibnamefont {Minohara}}, \bibinfo {author} {\bibfnamefont
  {H.}~\bibnamefont {Kumigashira}}, \bibinfo {author} {\bibfnamefont
  {Y.}~\bibnamefont {Yamasaki}}, \bibinfo {author} {\bibfnamefont
  {H.}~\bibnamefont {Nakao}}, \bibinfo {author} {\bibfnamefont
  {Y.}~\bibnamefont {Murakami}}, \bibinfo {author} {\bibfnamefont
  {A.}~\bibnamefont {Fujimori}}, \ and\ \bibinfo {author} {\bibfnamefont
  {H.}~\bibnamefont {Wadati}},\ }\href {\doibase 10.1103/PhysRevB.97.075134}
  {\bibfield  {journal} {\bibinfo  {journal} {Phys. Rev. B}\ }\textbf {\bibinfo
  {volume} {97}},\ \bibinfo {pages} {075134} (\bibinfo {year}
  {2018})}\BibitemShut {NoStop}%
\bibitem [{\citenamefont {Okamoto}\ \emph {et~al.}(2010)\citenamefont
  {Okamoto}, \citenamefont {Huang}, \citenamefont {Chao}, \citenamefont
  {Huang}, \citenamefont {Hsu}, \citenamefont {Fujimori}, \citenamefont
  {Masuno}, \citenamefont {Terashima}, \citenamefont {Takano},\ and\
  \citenamefont {Chen}}]{Okamoto2010}%
  \BibitemOpen
  \bibfield  {author} {\bibinfo {author} {\bibfnamefont {J.}~\bibnamefont
  {Okamoto}}, \bibinfo {author} {\bibfnamefont {D.~J.}\ \bibnamefont {Huang}},
  \bibinfo {author} {\bibfnamefont {K.~S.}\ \bibnamefont {Chao}}, \bibinfo
  {author} {\bibfnamefont {S.~W.}\ \bibnamefont {Huang}}, \bibinfo {author}
  {\bibfnamefont {C.-H.}\ \bibnamefont {Hsu}}, \bibinfo {author} {\bibfnamefont
  {A.}~\bibnamefont {Fujimori}}, \bibinfo {author} {\bibfnamefont
  {A.}~\bibnamefont {Masuno}}, \bibinfo {author} {\bibfnamefont
  {T.}~\bibnamefont {Terashima}}, \bibinfo {author} {\bibfnamefont
  {M.}~\bibnamefont {Takano}}, \ and\ \bibinfo {author} {\bibfnamefont {C.~T.}\
  \bibnamefont {Chen}},\ }\href {\doibase 10.1103/PhysRevB.82.132402}
  {\bibfield  {journal} {\bibinfo  {journal} {Phys. Rev. B}\ }\textbf {\bibinfo
  {volume} {82}},\ \bibinfo {pages} {132402} (\bibinfo {year}
  {2010})}\BibitemShut {NoStop}%
\bibitem [{\citenamefont {Minohara}\ \emph {et~al.}(2016)\citenamefont
  {Minohara}, \citenamefont {Kitamura}, \citenamefont {Wadati}, \citenamefont
  {Nakao}, \citenamefont {Kumai}, \citenamefont {Murakami},\ and\ \citenamefont
  {Kumigashira}}]{Minohara2016}%
  \BibitemOpen
  \bibfield  {author} {\bibinfo {author} {\bibfnamefont {M.}~\bibnamefont
  {Minohara}}, \bibinfo {author} {\bibfnamefont {M.}~\bibnamefont {Kitamura}},
  \bibinfo {author} {\bibfnamefont {H.}~\bibnamefont {Wadati}}, \bibinfo
  {author} {\bibfnamefont {H.}~\bibnamefont {Nakao}}, \bibinfo {author}
  {\bibfnamefont {R.}~\bibnamefont {Kumai}}, \bibinfo {author} {\bibfnamefont
  {Y.}~\bibnamefont {Murakami}}, \ and\ \bibinfo {author} {\bibfnamefont
  {H.}~\bibnamefont {Kumigashira}},\ }\href {\doibase 10.1063/1.4958670}
  {\bibfield  {journal} {\bibinfo  {journal} {J. Appl. Phys.}\ }\textbf
  {\bibinfo {volume} {120}},\ \bibinfo {pages} {025303} (\bibinfo {year}
  {2016})}\BibitemShut {NoStop}%
\bibitem [{\citenamefont {Holldack}\ \emph {et~al.}(2014)\citenamefont
  {Holldack}, \citenamefont {Bahrdt}, \citenamefont {Balzer}, \citenamefont
  {Bovensiepen}, \citenamefont {Brzhezinskaya}, \citenamefont {Erko},
  \citenamefont {Eschenlohr}, \citenamefont {Follath}, \citenamefont {Firsov},
  \citenamefont {Frentrup}, \citenamefont {{Le Guyader}}, \citenamefont
  {Kachel}, \citenamefont {Kuske}, \citenamefont {Mitzner}, \citenamefont
  {M{\"{u}}ller}, \citenamefont {Pontius}, \citenamefont {Quast}, \citenamefont
  {Radu}, \citenamefont {Schmidt}, \citenamefont
  {Sch{\"{u}}{\ss}ler-Langeheine}, \citenamefont {Sperling}, \citenamefont
  {Stamm}, \citenamefont {Trabant},\ and\ \citenamefont
  {F{\"{o}}hlisch}}]{Holldack2014}%
  \BibitemOpen
  \bibfield  {author} {\bibinfo {author} {\bibfnamefont {K.}~\bibnamefont
  {Holldack}}, \bibinfo {author} {\bibfnamefont {J.}~\bibnamefont {Bahrdt}},
  \bibinfo {author} {\bibfnamefont {A.}~\bibnamefont {Balzer}}, \bibinfo
  {author} {\bibfnamefont {U.}~\bibnamefont {Bovensiepen}}, \bibinfo {author}
  {\bibfnamefont {M.}~\bibnamefont {Brzhezinskaya}}, \bibinfo {author}
  {\bibfnamefont {A.}~\bibnamefont {Erko}}, \bibinfo {author} {\bibfnamefont
  {A.}~\bibnamefont {Eschenlohr}}, \bibinfo {author} {\bibfnamefont
  {R.}~\bibnamefont {Follath}}, \bibinfo {author} {\bibfnamefont
  {A.}~\bibnamefont {Firsov}}, \bibinfo {author} {\bibfnamefont
  {W.}~\bibnamefont {Frentrup}}, \bibinfo {author} {\bibfnamefont
  {L.}~\bibnamefont {{Le Guyader}}}, \bibinfo {author} {\bibfnamefont
  {T.}~\bibnamefont {Kachel}}, \bibinfo {author} {\bibfnamefont
  {P.}~\bibnamefont {Kuske}}, \bibinfo {author} {\bibfnamefont
  {R.}~\bibnamefont {Mitzner}}, \bibinfo {author} {\bibfnamefont
  {R.}~\bibnamefont {M{\"{u}}ller}}, \bibinfo {author} {\bibfnamefont
  {N.}~\bibnamefont {Pontius}}, \bibinfo {author} {\bibfnamefont
  {T.}~\bibnamefont {Quast}}, \bibinfo {author} {\bibfnamefont
  {I.}~\bibnamefont {Radu}}, \bibinfo {author} {\bibfnamefont {J.-S.}\
  \bibnamefont {Schmidt}}, \bibinfo {author} {\bibfnamefont {C.}~\bibnamefont
  {Sch{\"{u}}{\ss}ler-Langeheine}}, \bibinfo {author} {\bibfnamefont
  {M.}~\bibnamefont {Sperling}}, \bibinfo {author} {\bibfnamefont
  {C.}~\bibnamefont {Stamm}}, \bibinfo {author} {\bibfnamefont
  {C.}~\bibnamefont {Trabant}}, \ and\ \bibinfo {author} {\bibfnamefont
  {A.}~\bibnamefont {F{\"{o}}hlisch}},\ }\href {\doibase
  10.1107/S1600577514012247} {\bibfield  {journal} {\bibinfo  {journal} {J.
  Synchrotron Radiat.}\ }\textbf {\bibinfo {volume} {21}},\ \bibinfo {pages}
  {1090} (\bibinfo {year} {2014})}\BibitemShut {NoStop}%
\bibitem [{\citenamefont {Nakao}\ \emph {et~al.}(2014)\citenamefont {Nakao},
  \citenamefont {Yamasaki}, \citenamefont {Okamoto}, \citenamefont {Sudayama},
  \citenamefont {Takahashi}, \citenamefont {Kobayashi}, \citenamefont {Kumai},\
  and\ \citenamefont {Murakami}}]{Nakao2014}%
  \BibitemOpen
  \bibfield  {author} {\bibinfo {author} {\bibfnamefont {H.}~\bibnamefont
  {Nakao}}, \bibinfo {author} {\bibfnamefont {Y.}~\bibnamefont {Yamasaki}},
  \bibinfo {author} {\bibfnamefont {J.}~\bibnamefont {Okamoto}}, \bibinfo
  {author} {\bibfnamefont {T.}~\bibnamefont {Sudayama}}, \bibinfo {author}
  {\bibfnamefont {Y.}~\bibnamefont {Takahashi}}, \bibinfo {author}
  {\bibfnamefont {K.}~\bibnamefont {Kobayashi}}, \bibinfo {author}
  {\bibfnamefont {R.}~\bibnamefont {Kumai}}, \ and\ \bibinfo {author}
  {\bibfnamefont {Y.}~\bibnamefont {Murakami}},\ }\href {\doibase
  10.1088/1742-6596/502/1/012015} {\bibfield  {journal} {\bibinfo  {journal}
  {J. Phys. Conf. Ser.}\ }\textbf {\bibinfo {volume} {502}},\ \bibinfo {pages}
  {012015} (\bibinfo {year} {2014})}\BibitemShut {NoStop}%
\bibitem [{\citenamefont {MacChesney}\ \emph {et~al.}(1965)\citenamefont
  {MacChesney}, \citenamefont {Sherwood},\ and\ \citenamefont
  {Potter}}]{MacChesney1965}%
  \BibitemOpen
  \bibfield  {author} {\bibinfo {author} {\bibfnamefont {J.~B.}\ \bibnamefont
  {MacChesney}}, \bibinfo {author} {\bibfnamefont {R.~C.}\ \bibnamefont
  {Sherwood}}, \ and\ \bibinfo {author} {\bibfnamefont {J.~F.}\ \bibnamefont
  {Potter}},\ }\href {\doibase 10.1063/1.1697052} {\bibfield  {journal}
  {\bibinfo  {journal} {J. Chem. Phys.}\ }\textbf {\bibinfo {volume} {43}},\
  \bibinfo {pages} {1907} (\bibinfo {year} {1965})}\BibitemShut {NoStop}%
\bibitem [{\citenamefont {Reehuis}\ \emph {et~al.}(2012)\citenamefont
  {Reehuis}, \citenamefont {Ulrich}, \citenamefont {Maljuk}, \citenamefont
  {Niedermayer}, \citenamefont {Ouladdiaf}, \citenamefont {Hoser},
  \citenamefont {Hofmann},\ and\ \citenamefont {Keimer}}]{Reehuis2012}%
  \BibitemOpen
  \bibfield  {author} {\bibinfo {author} {\bibfnamefont {M.}~\bibnamefont
  {Reehuis}}, \bibinfo {author} {\bibfnamefont {C.}~\bibnamefont {Ulrich}},
  \bibinfo {author} {\bibfnamefont {A.}~\bibnamefont {Maljuk}}, \bibinfo
  {author} {\bibfnamefont {C.}~\bibnamefont {Niedermayer}}, \bibinfo {author}
  {\bibfnamefont {B.}~\bibnamefont {Ouladdiaf}}, \bibinfo {author}
  {\bibfnamefont {A.}~\bibnamefont {Hoser}}, \bibinfo {author} {\bibfnamefont
  {T.}~\bibnamefont {Hofmann}}, \ and\ \bibinfo {author} {\bibfnamefont
  {B.}~\bibnamefont {Keimer}},\ }\href {\doibase 10.1103/PhysRevB.85.184109}
  {\bibfield  {journal} {\bibinfo  {journal} {Phys. Rev. B}\ }\textbf {\bibinfo
  {volume} {85}},\ \bibinfo {pages} {184109} (\bibinfo {year}
  {2012})}\BibitemShut {NoStop}%
\bibitem [{\citenamefont {Ishikawa}\ \emph {et~al.}(1998)\citenamefont
  {Ishikawa}, \citenamefont {Park}, \citenamefont {Katsufuji}, \citenamefont
  {Arima},\ and\ \citenamefont {Tokura}}]{Ishikawa1998}%
  \BibitemOpen
  \bibfield  {author} {\bibinfo {author} {\bibfnamefont {T.}~\bibnamefont
  {Ishikawa}}, \bibinfo {author} {\bibfnamefont {S.~K.}\ \bibnamefont {Park}},
  \bibinfo {author} {\bibfnamefont {T.}~\bibnamefont {Katsufuji}}, \bibinfo
  {author} {\bibfnamefont {T.}~\bibnamefont {Arima}}, \ and\ \bibinfo {author}
  {\bibfnamefont {Y.}~\bibnamefont {Tokura}},\ }\href {\doibase
  10.1103/PhysRevB.58.R13326} {\bibfield  {journal} {\bibinfo  {journal} {Phys.
  Rev. B}\ }\textbf {\bibinfo {volume} {58}},\ \bibinfo {pages} {R13326}
  (\bibinfo {year} {1998})}\BibitemShut {NoStop}%
\bibitem [{\citenamefont {Zhu}\ \emph {et~al.}(2018)\citenamefont {Zhu},
  \citenamefont {Hoffman}, \citenamefont {Rowland}, \citenamefont {Park},
  \citenamefont {Walko}, \citenamefont {Freeland}, \citenamefont {Ryan},
  \citenamefont {Schaller}, \citenamefont {Bhattacharya},\ and\ \citenamefont
  {Wen}}]{Zhu2018}%
  \BibitemOpen
  \bibfield  {author} {\bibinfo {author} {\bibfnamefont {Y.}~\bibnamefont
  {Zhu}}, \bibinfo {author} {\bibfnamefont {J.}~\bibnamefont {Hoffman}},
  \bibinfo {author} {\bibfnamefont {C.~E.}\ \bibnamefont {Rowland}}, \bibinfo
  {author} {\bibfnamefont {H.}~\bibnamefont {Park}}, \bibinfo {author}
  {\bibfnamefont {D.~A.}\ \bibnamefont {Walko}}, \bibinfo {author}
  {\bibfnamefont {J.~W.}\ \bibnamefont {Freeland}}, \bibinfo {author}
  {\bibfnamefont {P.~J.}\ \bibnamefont {Ryan}}, \bibinfo {author}
  {\bibfnamefont {R.~D.}\ \bibnamefont {Schaller}}, \bibinfo {author}
  {\bibfnamefont {A.}~\bibnamefont {Bhattacharya}}, \ and\ \bibinfo {author}
  {\bibfnamefont {H.}~\bibnamefont {Wen}},\ }\href {\doibase
  10.1038/s41467-018-04199-4} {\bibfield  {journal} {\bibinfo  {journal} {Nat.
  Commun.}\ }\textbf {\bibinfo {volume} {9}},\ \bibinfo {pages} {1799}
  (\bibinfo {year} {2018})}\BibitemShut {NoStop}%
\bibitem [{\citenamefont {Fujioka}\ \emph {et~al.}(2012)\citenamefont
  {Fujioka}, \citenamefont {Ishiwata}, \citenamefont {Kaneko}, \citenamefont
  {Taguchi},\ and\ \citenamefont {Tokura}}]{Fujioka2012}%
  \BibitemOpen
  \bibfield  {author} {\bibinfo {author} {\bibfnamefont {J.}~\bibnamefont
  {Fujioka}}, \bibinfo {author} {\bibfnamefont {S.}~\bibnamefont {Ishiwata}},
  \bibinfo {author} {\bibfnamefont {Y.}~\bibnamefont {Kaneko}}, \bibinfo
  {author} {\bibfnamefont {Y.}~\bibnamefont {Taguchi}}, \ and\ \bibinfo
  {author} {\bibfnamefont {Y.}~\bibnamefont {Tokura}},\ }\href {\doibase
  10.1103/PhysRevB.85.155141} {\bibfield  {journal} {\bibinfo  {journal} {Phys.
  Rev. B}\ }\textbf {\bibinfo {volume} {85}},\ \bibinfo {pages} {155141}
  (\bibinfo {year} {2012})}\BibitemShut {NoStop}%
\end{thebibliography}%
\end{document}